\documentclass{aa}
\usepackage{epsfig}
\usepackage{graphicx}
\usepackage{psfig,epsf}

\def\beq{\begin{equation}}
\def\eeq{\end{equation}}

\def\cm3{cm$^{-3}$}

\begin{document}

\title{Distance determinations using type II supernovae and the expanding photosphere method}
\subtitle{}

\author{Luc Dessart\inst{1,3}
        \and
        D. John Hillier\inst{2}
        }
\offprints{Luc Dessart,\\ \email{luc@as.arizona.edu}}

  \institute{Max-Planck-Institut f\"{u}r Astrophysik,
             Karl-Schwarzschild-Str 1, 85748, Garching bei M\"{u}nchen, Germany 
        \and
           Department of Physics and Astronomy, University of Pittsburgh,
           3941 O'Hara Street, Pittsburgh, PA, 15260 
         \and
           Steward Observatory, University of Arizona, 933 North Cherry Avenue,
        Tucson, AZ 85721, USA
            }

\date{Accepted/Received}

\abstract{
Due to their high intrinsic brightness, caused by the disruption of the progenitor envelope 
by the shock-wave initiated at the bounce of the collapsing core, 
hydrogen-rich (type II) supernovae (SN) can be used as lighthouses
to constrain distances in the Universe using variants of the Baade-Wesselink method.
Based on a large set of CMFGEN models (Hillier \& Miller 1998) covering the 
photospheric phase of type II SN, we study the various concepts entering 
one such technique, the Expanding Photosphere Method (EPM).

We compute correction factors $\xi$ needed to approximate the synthetic Spectral 
Energy Distribution (SED) with that of a blackbody at temperature $T$.
Our $\xi$, although similar, are systematically greater, by $\sim 0.1$, than the
values obtained by Eastman et al. (1996) and 
translate into a systematic enhancement of 10-20\% in EPM-distances.
We find that line emission and absorption, not directly linked to color temperature 
variations, can considerably alter the synthetic magnitude: in particular,
line-blanketing attributable to Fe{\,\sc ii} and Ti{\,\sc ii} is the principal cause
for above-unity correction factors in the $B$ and $V$ bands in hydrogen-recombining 
models.


Following the dominance of electron-scattering opacity in type II SN outflows, the 
blackbody SED arising at the thermalization depth is diluted, by a factor of approximately 
0.2 to 0.4
for fully- or partially-ionized models, but rising to unity as hydrogen recombines 
for effective temperatures below 9,000\,K.
For a given effective temperature, models with a larger spatial scale,
or lower density exponent, have a larger electron-scattering optical
depth at the photosphere and consequently suffer enhanced dilution.
We also find that when lines are present in the emergent spectrum, the photospheric 
radius in the corresponding wavelength range can be enhanced by
a factor of 2-3 compared to the case when only continuum opacity is considered.
Lines can thus nullify the uniqueness of the photosphere radius and invalidate the 
Baade method at the heart of the EPM.
Both the impact of line-blanketing on the SED and the photospheric radius at low $T$
suggest that the EPM is best used at early times, when the outflow is fully ionized
and line-opacity mostly confined to the UV range.

We also investigate how reliably one can measure the photospheric
velocity from P-Cygni line profiles. Contrary to the usually held belief,
the velocity at maximum absorption in the P-Cygni trough of
optically-thick lines can both {\it overestimate or underestimate} the
photospheric velocity, with a magnitude that depends on the 
SN outflow density gradient and the optical thickness of the line. 
This stems from the wavelength-shift, toward line-center, of the 
location of maximum line-absorption for rays with larger impact parameters.
This has implications for the measurement of expansion rates in SN outflows, 
especially at earlier times when only fewer, broader, and blue-shifted lines 
are present. This investigation should facilitate more reliable use of the
EPM and the determination of distances in the Universe using type II SN.

\keywords{radiative transfer -- stars: atmospheres -- stars:
supernovae -- line: formation
          }
}
\titlerunning{type II SN as distance indicators}
\maketitle

\section{Introduction}
\label{Sec_intro}

In Dessart \& Hillier (2005a, hereafter Paper I), we have presented
first and general results obtained with the non-LTE model atmosphere
code CMFGEN (Hillier \& Miller 1998) for the quantitative spectroscopic
analysis of photospheric-phase type II supernovae (SN).
We demonstrated the ability of CMFGEN to reproduce, with great accuracy,
the evolution of type II SN spectra.
The only noticeable difficulty experienced by the code is in fitting
H$\alpha$ during the hydrogen-recombination phase, which is underestimated
both in absorption and emission by up to a factor of 2-3 compared to
observations.
However, CMFGEN does a superb job for earlier times, both for the continuum
and the lines, and from UV to near-IR wavelengths.
This gives us confidence that CMFGEN can provide information, at a very
competitive level, on the properties of the SN ejecta and
those of the progenitor.

In this paper, we leave this aspect aside and focus on using CMFGEN to discuss
current methods that use type II SN for distance determinations in the Universe.
Being typically a factor of 10,000 times as bright as Cepheids, type II SN 
constitute an excellent tool to set distances over a larger volume of the Universe, 
while providing an alternative to type Ia SN and their assumed uniform 
properties.

Two methods are used to determine distances in the Universe based
on type II SN: the Expanding Photosphere Method (EPM; Kirshner \& Kwan 1974,
Eastman \& Kirshner 1989) and the Spectral-fitting Expanding Atmosphere
Method (SEAM; Baron et al. 1995, 2004).
The EPM is based on the Baade method (Baade 1926), which requires
a unique and well-defined radius for the expanding photosphere.
As we shall see below, this assumption is only valid at early times when
the outflow is fully ionized, i.e. when effects due to metal line-blanketing 
in the optical are small.
While the EPM approximates the Spectral Energy Distribution (SED) with a blackbody, adjusting 
for inconsistencies by means of correction factors (Eastman et al. 1996, hereafter E96; 
see Sect.~\ref{Sec_pres_epm}-\ref{Sec_corr_fac}), the SEAM uses synthetic spectra to 
directly fit the observed SED, and thus bypasses the use of correction factors.

Distance determinations, based on the EPM, have shown different degrees
of consistency with the Cepheid-based distance.
The EPM works fine for the LMC (using SN1987A, Eastman \& Kirshner 1989) but
predicts a distance to NGC\,1637 50\% smaller (using SN1999em,
Leonard et al. 2002) than found using Cepheids (Leonard et al. 2003).
On the contrary, every time the SEAM has been used, it has delivered the 
previously determined Cepheid distance.
Why does the EPM work fine when applied to SN1987A and not for SN1999em?
Indeed, the EPM should give an accurate distance, provided one uses the 
appropriate correction factors and photospheric velocity. Thus of crucial 
importance to the EPM method is an understanding of the errors and biases 
associated with the derived correction factors and photospheric velocities.

Due to its ease of use, and since it does not require
flux calibrated spectra, it is desirable to investigate whether the EPM
technique could be improved, and thus used reliably for distance determinations.
In this paper we devote our attention to the EPM and the
various elements involved that may introduce errors into distance
determinations: most importantly the correction factors, and the photospheric velocity
measured from P-Cygni line profiles.

This paper is structured as follows.
In Sect.~\ref{Sec_pres_epm}, we present the various aspects entering the EPM.
In Sect.~\ref{Sec_corr_fac}, we study what {\it correction} factors we obtain
in order to approximate the SED of type II SN with a blackbody.
We also compare with equivalent values presented in E96. 
In Sect.~\ref{Sec_dil_fac}, we investigate the physical origin and
properties of flux dilution by electron-scattering. 
We focus on one key quantity controlling the magnitude of 
such a dilution, i.e. the ratio of thermalization to photospheric radius.
In Sect.~\ref{Sec_vphot}, we discuss how the model photospheric velocity
compares with various measurements performed on P-Cygni line profiles in the
optical range.
Combining these various insights, we conclude in Sect.~\ref{Sec_conc} by
presenting the prerequisite for minimizing errors in the EPM and the
Baade-Wesselink methods.

\section{Presentation of the EPM method}
\label{Sec_pres_epm}

  Developed for spatially-resolved outbursts, the Baade method (1926) 
constrains the distance $D$ to a novae based on measures of the rate of 
expansion $v_{\rm phot}$ of the radiating photospheric surface of 
radius $R_{\rm phot}$, and of its apparent angular size $\theta \equiv R_{\rm phot}/D$, following 
the formula,

\beq
D = v_{\rm phot} \Delta t / \theta
\, ,
\label{eq1}
\eeq

\noindent
where $\Delta t$ represents the difference between the time of observation $t$
and that of outburst $t_0$.
Here we have made the key assumption of homologous expansion, which stipulates that 
mass shells, after a prompt acceleration at $t_0$, travel at constant
velocity thereafter. 
We further neglect the original radius of the object, typically much smaller 
than corresponding values at times $t$ of the order of days.
When the outburst date $t_0$ is not known, one uses a sample of observations to
fit the distribution of points along the line defined by
$t = (\theta/v_{\rm phot})D + t_0$, with slope $D$ and ordinate crossing $t_0$.

In principle, this method is directly applicable to a SN explosion suitably positioned
within the Milky Way.
These events are, however, too rare, and daily SN discoveries are instead permitted
by searching over tens-of-thousands of galaxies.
Such SN are too far to be resolved spatially, and thus, their angular size $\theta$ must
be obtained indirectly.

For type II SN, the EPM estimates $\theta$ by assuming that the outflow 
radiates as a blackbody at temperature $T$, yielding the relation,
\beq
    4 \pi R_{\rm phot}^2 \pi B_{\nu}(T) = 4 \pi D^2 f^{\rm dered}_\nu
\, ,
\label{eq2}
\eeq
where all quantities (omitting $D$) are evaluated at frequency $\nu$:
$f^{\rm dered}_\nu$ is the de-reddened observed flux, i.e.
$f^{\rm dered}_\nu = 10^{0.4 A_\nu} \, f^{\rm obs}_{\nu}$,
with $A_{\nu}$ the extinction and $f^{\rm obs}_{\nu}$ the observed flux.

Equation (2), however, is ambiguous: the blackbody temperature $T$
corresponds to the outflow layer where photons are thermalized ($R_{\rm th}$), 
rather than the layer from where they escape ($R_{\rm phot}$), defined as the location 
where the total inward-integrated radial optical depth equals 2/3.
At early times of evolution, when the SN outflow is fully- or partially-ionized,
$R_{\rm th} < R_{\rm phot}$ because the electron-scattering opacity 
contributes significantly to the total opacity at the photosphere.
Although this gray-opacity preserves the relative wavelength distribution of the 
Planck function $\pi B_{\nu}(T)$, set at $R_{\rm th}$, it introduces a global flux 
dilution.

Schmutz et al. (1990) argued that, due to the finite extent of the photosphere, 
$R_{\rm th}$ covers a sizable range of radii, increasing toward longer wavelengths,
thereby associated with decreasing thermal temperatures and weaker dilution. 
When combined, these effects produce an emergent flux distribution characterized by
a cooler blackbody temperature than that found at $R_{\rm th}$.
Thus, to account for ``sphericity'' and dilution effects in Eq.~(\ref{eq2}), 
we replace $T$ by the 
color temperature $T_c$, and $R_{\rm phot}$ with $\xi R_{\rm phot}$, with explanations on
how to obtain $T_c$ and $\xi$ given below. A more thorough discussion
of the origin of the correction factor, $\xi$, is given in Sect.~\ref{Sec_dil_fac}.

Taking these considerations into account
 , we can express the observed flux as,
\beq
   f^{\rm obs}_\nu = \xi^2 \, \theta^2 \, 10^{-0.4 A_\nu} \, \pi B_\nu(T_c)
\, .
\label{eq3}
\eeq

The quantity $\xi$, often called dilution factor (e.g., E96; Hamuy et al. 2001, H01), 
represents in fact a general correction to be applied to the fitted 
blackbody distribution in order to reproduce the observed flux at $\nu$.


By integrating each side of Eq.~(\ref{eq3}) over the transmission function $\phi_{\bar{\nu},\nu}$ 
of filter $\bar{\nu}$, we now obtain an equation relating magnitudes,

\beq
  m_{\bar{\nu}} = -5 \log(\xi) -5 \log(\theta) + A_{\bar{\nu}} + b_{\bar{\nu}}
\, ,
\label{eq4}
\eeq
where
\beq
b_{\bar{\nu}} = -2.5 \log\left( \int_0^{\infty} d\nu \, \phi_{\bar{\nu},\nu} \, 
\pi B_{\nu}(T_c)  \right) + C_{\bar{\nu}}
\, .
\label{eq5}
\eeq
Here, we adopt the transmission functions $\phi_{\bar{\nu},\nu}$ and zero-point constants
$C_{\bar{\nu}}$ from H01.

In practice, the color temperature $T_c$ is obtained by matching the slope of the corresponding
blackbody to a set of magnitudes, usually a combination $S$ taken from the optical 
and near-IR band passes $UBVRIJHK$.
Not all band passes are equally suitable for use in the EPM: 
the fast-decreasing flux in the $U$-band makes the corresponding magnitude 
quickly too faint for good-quality observation, so the $U$-band is usually 
discarded; the $R$-band also has problems due to its overlap with the H$\alpha$ 
line, which shows significant variation in strength and width amongst 
type II SN.
To allow a complete comparison between our results and those of E96, we 
sometimes quote in this work the results obtained for the near-IR band 
passes, although such magnitudes are rarely available.
The various combinations $S$ usually adopted are thus, $\{B,V\}$, 
$\{B,V,I\}$, $\{V,I\}$, and $\{J,H,K\}$.

The procedure is then, for a given set of photometric observations at time $t$,
to apply a minimization of the quantity $\epsilon$ defined as
\beq
  \epsilon = \sum_{\bar{\nu} \in S} \left[
m_{\bar{\nu}} + 5 \log [\theta\xi_S] - A_{\bar{\nu}} - b_{\bar{\nu}}(T_S)
\right]^2
\, .
\label{eq6}
\eeq
For a given bandpass combination $S$ and corresponding magnitudes, one
obtains an $S$-dependent color temperature $T_S$: we thus write $\theta$ as 
$\theta_S$ and $\xi$ as $\xi_S$.

For each date, one can thus determine the values of $\theta_S$ and $\xi_S$
for all possible combinations of filters $S$ as well as the corresponding value
of $v_{\rm phot}$ from spectroscopic measurements.
Grouping observations taken over a number of nights, one can then plot the quantity,
\beq
    t = D \frac{\theta_S}{v_{\rm phot}} + t_0
\, .
\label{eq7}
\eeq
Fitting a line to the resulting distribution of points delivers
the distance D (the slope) and the time of explosion (the ordinate
crossing).
One can compare the results for different choices of band passes $S$.

%
%

A fundamental part in the EPM is to quantify, in Eq.~(\ref{eq6}),
the correction factor $\xi_S$ associated with the blackbody color 
temperature $T_S$, for each bandpass set $S$.
The procedure is to recast Eq.~(\ref{eq4}) in terms of absolute magnitude
and distance modulus,

\beq
M_{\bar{\nu}} = -5 \log \xi_S -5 \log \left(\frac{R_{\rm phot}}{10 \rm pc} \right)
                + b_{\bar{\nu}}(T_S)
\, .
\label{eq8}
\eeq

Then, based on a large set of model atmosphere calculations providing measures of
$R_{\rm phot}$ and $M_{\bar{\nu}}$,  one can minimize, for the different bandpass 
choices $S$, the quantity $\epsilon$ given by,
 
\beq
  \epsilon = \sum_{\bar{\nu} \in S} \left[
M_{\bar{\nu}} + 5 \log [\xi_S] +5 \log \left(\frac{R_{\rm phot}}{10 \rm pc}
\right) - b_{\bar{\nu}}(T_S)
\right]^2
\, .
\label{eq9}
\eeq

E96 provides analytical formulations to describe how $\xi_S$ changes with $T_S$, 
subsequently revised by H01 to include a more suitable set
of filter transmission functions and zero-points.
Such formulations have been used by observers to obtain distance estimates
to type II SN (see, e.g., Schmidt et al. 1994ab; E96; H01; L02).
In Sect.~\ref{Sec_corr_fac}, based on CMFGEN models, we compute similar relations to
compare and gauge the uncertainties, since these translate directly into the EPM-distance.
In Sect.~\ref{Sec_dil_fac}, we make a digression on the physical origin of flux dilution,
its properties for a wide variety of type II SN and its correspondence with correction factors.
In Sect.~\ref{Sec_vphot}, we then discuss the procedure to infer the photospheric velocity,
a quantity whose uncertainty weighs on the EPM just as much as correction factors do.


\section{Correction factors}
\label{Sec_corr_fac}

 The EPM-distance, obtained by minimizing $\epsilon$ (Eq.~\ref{eq6}), requires knowledge
of the dependence of correction factors $\xi_S$ on blackbody color temperature
$T_S$, whose analytical description, at present, is only given in the work of E96.
Here, following E96 and the casting of $\epsilon$ given in Eq.~(\ref{eq9}), we compute the 
variation of $\xi_S$ with $T_S$ based on a large number of CMFGEN models that cover 
the type II SN evolution over the first few weeks after core-collapse; then, we compare 
with the results of E96.

\begin{figure}[htp!]
\hspace{-1.5cm} \epsfig{file=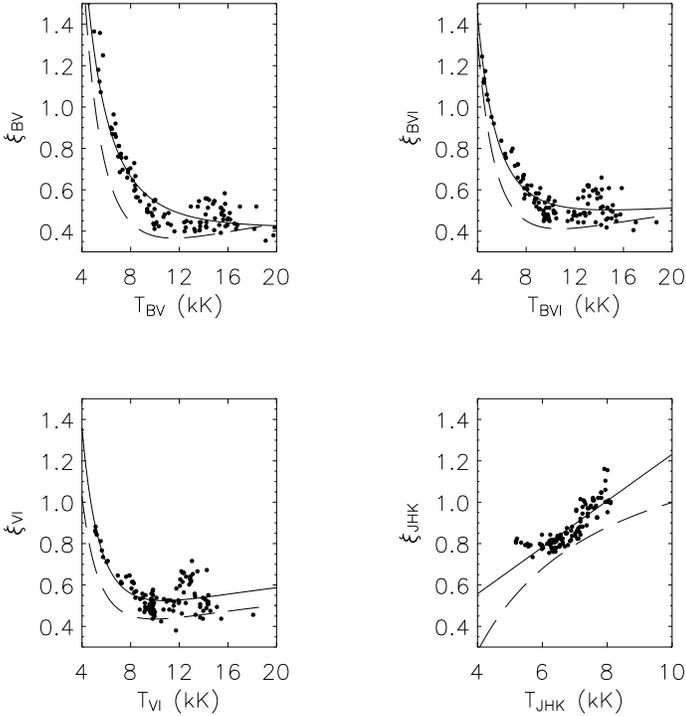, width=10.5cm}
\caption{
Correction factors $\xi_S$ versus the corresponding blackbody color
temperature $T_S$, obtained simultaneously by minimizing $\epsilon$ (Eq.~\ref{eq9}).
We show solutions for four different bandpass combinations,
$\{B,V\}$, $\{B,V,I\}$, $\{V,I\}$, and $\{J,H,K\}$.
In each panel, we overplot a first- or second-order polynomial of the form
$\xi_{\rm fit} = \sum_i a_i (10^4 {\rm K}/T_S)^i$, with coefficients $a_i$
obtained from a best-fit to these data points 
(solid line; see Table~\ref{tab_xi}) or adopted from H01 (dashed line).
For ease of comparison, a unique ordinate range is used for all plots.
}
\label{fig_xis_ts}
\end{figure}

\begin{figure}[htp!]
\hspace{-1.2cm} \epsfig{file=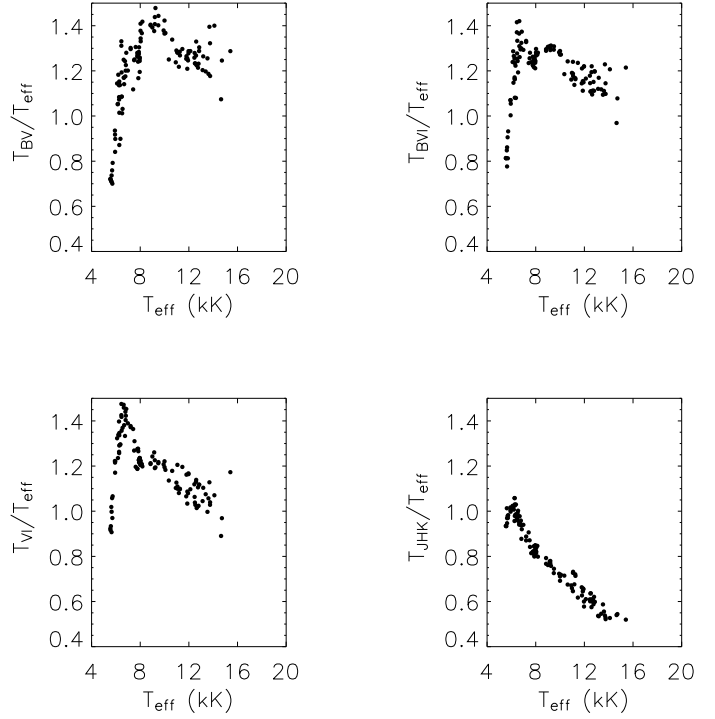, width=10.5cm}
\caption{
Solutions obtained from Eq.~(\ref{eq9}), this time showing the color temperature
in each bandpass combination $S$ normalized by, and plotted as a function 
of, $T_{\rm eff}$. The band passes are arranged as in Fig.~1, and for ease of 
comparison, a unique ordinate range is used for all plots.
}
\label{fig_ts_teff}
\end{figure}

\begin{figure}[htp!]
\hspace{-1.5cm} \epsfig{file=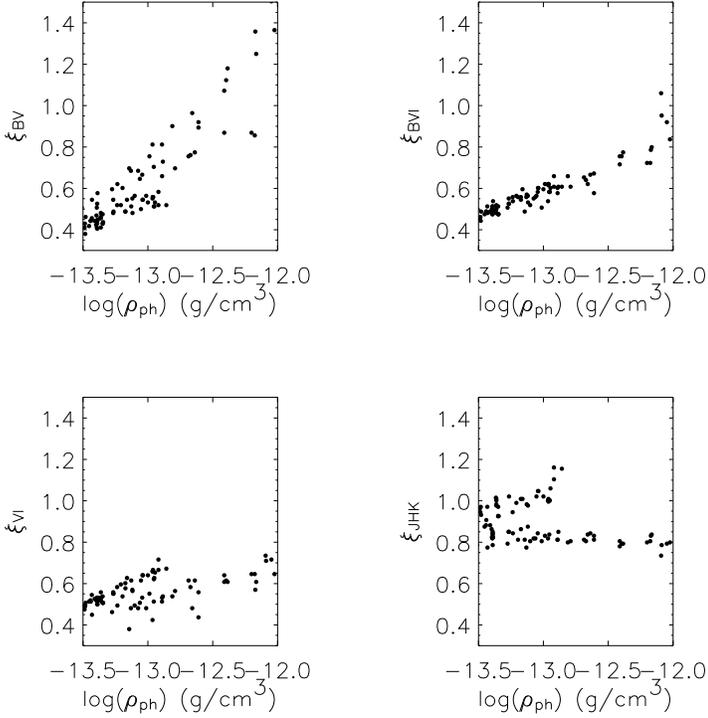, width=10.5cm}
\caption{
Same as Fig.~\ref{fig_xis_ts} showing the correction factors
as a function of photospheric density, again for each bandpass
combination $S$.
For ease of comparison, a unique ordinate range is used for all plots.
}
\label{fig_xis_rho}
\end{figure}

In Figs.~\ref{fig_xis_ts}, \ref{fig_ts_teff} and \ref{fig_xis_rho}, we plot various
combinations of the resulting quantities that minimize this last form of
$\epsilon$.
Figure~\ref{fig_xis_ts} shows the variation of the correction factors $\xi_S$
versus the blackbody color temperature, $T_S$, obtained for each of the four bandpass
combinations, $\{B,V\}$, $\{B,V,I\}$, $\{V,I\}$, and $\{J,H,K\}$.
Figure~\ref{fig_ts_teff} shows the variation of $T_S/T_{\rm eff}$ versus the
effective temperature $T_{\rm eff}$, defined here as the gas (electron) temperature
at continuum optical depth two-third (integrated inwards from the outer grid radius).
This location corresponds to the photospheric radius $R_{\rm phot}$.
Finally, Fig.~\ref{fig_xis_rho} shows the variation of $\xi_S$ with photospheric 
density $\rho_{\rm ph}$.
We refer the reader to E96 for a discussion of the properties shown in
Fig.~\ref{fig_ts_teff} and \ref{fig_xis_rho} (which are similar to theirs), and
here focus our attention to the relation $\xi_S$-$T_S$ illustrated in 
Fig.~\ref{fig_xis_ts}.

Let us focus on the top-left panel of Fig.~\ref{fig_xis_ts},
showing the variation of $\xi_{BV}$ as a function of blackbody 
color temperature $T_{BV}$.
The distribution of data points follows two distinct patterns:
above ca. 9,000\,K, $\xi_{BV}$ is roughly independent of $T_{BV}$ and 
of order 0.45, but below ca. 9,000\,K, it increases for lower values of $T_{BV}$, 
culminating at 1.5 for $T_{BV} \sim$ 4000\,K.
For a fixed temperature $T_{BV}$, the scatter is uniform and of order 0.1,  
corresponding to decreasing uncertainties towards lower temperature
(from 20 down to 10\%).
But over a range of temperature, this scatter increases, at 
low $T_{BV}$, by a factor 2-3, following the sharp rise of $\xi_{BV}$.
%
Thus, adopting a unique correction factor at a given temperature can 
introduce non-negligible errors, either through the neglect of the 
intrinsic scatter of $\xi_{BV}$, or through an inaccuracy in the temperature 
$T_{BV}$, made worse for smaller values of $T_{BV}$.
In addition, we overplot a polynomial with coefficients obtained from a
best-fit to our distribution of data points (solid line; 
see Table~\ref{tab_xi}), or extracted from Table~14 in H01 (dashed line).
Despite a qualitative agreement, our best-fit polynomial 
lies systematically above H01's or E96's, with a maximum difference 
of ca. 0.2 (or 50\%) at $T_{BV} \sim$ 8,000\,K.

Turning to the other panels in Fig.~\ref{fig_xis_ts}, we see that 
$\xi_{BVI}$ and $\xi_{VI}$ follow a similar trend, although 
$\xi_{VI}$ remains below unity at low temperatures $T_{VI}$.
For $\xi_{JHK}$, the trend is altogether different, correction factors 
decreasing at lower temperatures, with a scatter reduced compared to the 
previous three cases.
Note that the $\{J,H,K\}$ bandpass is poorly sensitive to color temperature 
in the 10,000\,K temperature range.
Here again, the overplotted polynomials show that our data points lie, 
on average, above those obtained by H01/E96.

The reason for the discrepancy is unclear.
It might be related to the different approach E96 adopts to 
handle the relativistic terms.
Also, E96 only solved the non-LTE problem for a few species, obtaining a consistent
distribution between absorptive and scattering opacity contributions, while for 
the rest, usually metals, line opacity was taken as pure scattering.
It is not clear that this can cause a sizable relative shift of the 
thermalization and photospheric radii, controlled by hydrogen bound-free and 
electron-scattering opacities (both handled in a similar way in E96 and here).
Finally, E96 handles metal line opacity with the ``expansion opacity'' formalism of 
Eastman \& Pinto (1993), a convenient but approximate approach compared to
the more consistent treatment adopted in CMFGEN, where all line opacity is
included systematically in a  ``brutal'' but accurate way.    
The only other obvious difference in our approach is the use of a very large model 
atom (manageable through the use of Super-Levels),
to allow a detailed and complete census of opacity sources, as well as
the non-LTE treatment of {\it all} species. 
These resulting differences are not academic because our systematically larger 
correction factors translate directly into EPM distances that are, on average, 
10-20\% larger than predicted by H01/E96 (Dessart \& Hillier 2005c).

\begin{table}
\caption[]{
Coefficients of the polynomial fits (shown as solid lines in Fig.~\ref{fig_xis_ts})
to the distribution of correction factors $\xi_S$ versus blackbody color temperature 
for the four bandpass combinations.
As in H01/E96, we use  a second-order polynomial of the form 
$\xi_{\rm fit} = \sum_i a_i (10^4 {\rm K}/T_S)^i$.
}
\begin{center}
\begin{tabular}{crrrc}
\hline
\hline
         &  \multicolumn{1}{c}{$\{B,V\}$}   &  $\{B,V,I\}$ &  \multicolumn{1}{c}{$\{V,I\}$} & $\{J,H,K\}$  \\
\hline
$a_0$    &     0.47188   &   0.63241   &    0.81662     &   0.10786    \\
$a_1$    &    -0.25399   &  -0.38375   &   -0.62896     &   1.12374     \\
$a_2$    &     0.32630   &   0.28425   &    0.33852     &   0.00000     \\
\hline
\end{tabular}
\end{center}
\label{tab_xi}
\end{table}
%

\begin{figure}[htp!]
\epsfig{file=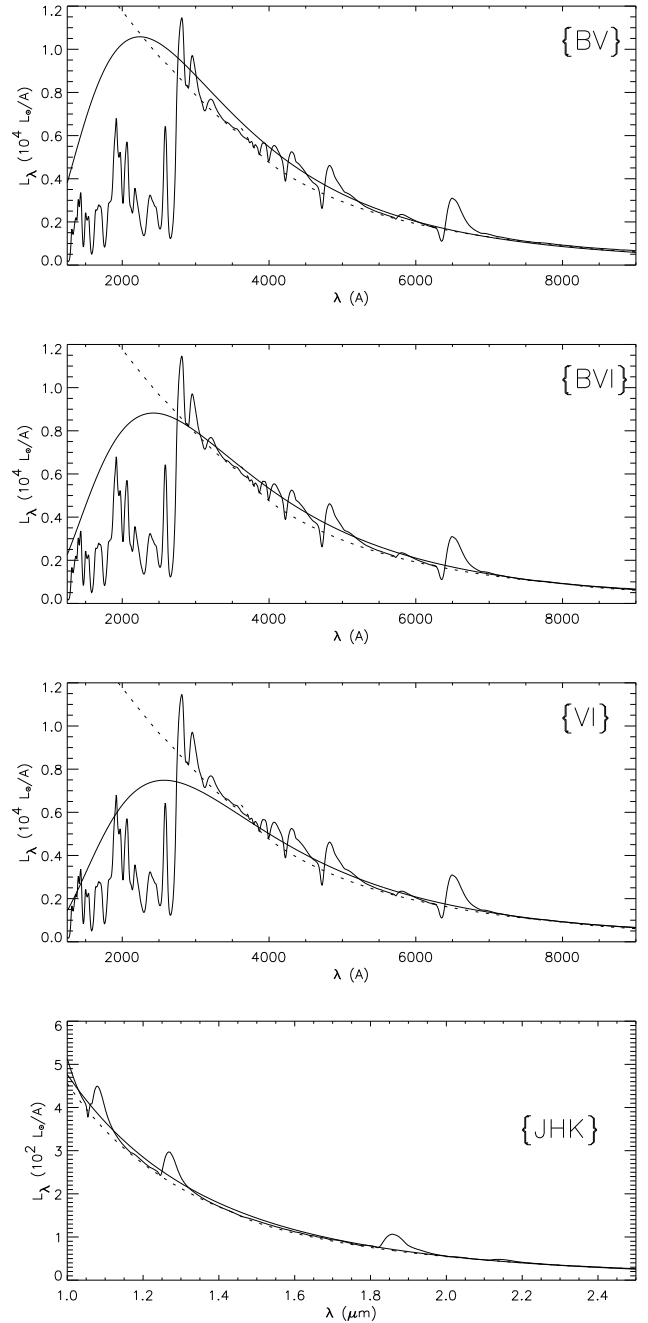, width=9cm}
\caption{
%
%
Comparison between the model with $T_{\rm eff}=$9,200\,K that fits well the observations of SN1999em
on the 5th of November 1999 (Sect. 4) with blackbody spectra whose temperature
is obtained by fitting the model magnitude in different band passes $\{B,V\}$,
$\{B,V,I\}$, $\{V,I\}$, and $\{J,H,K\}$, shown in this order from the top to the bottom panels.
The corresponding color temperatures and correction factors are:
12,945\,K  and 0.443 ($BV$), 11,951\,K  and 0.494 ($BVI$),
11,278\,K and 0.526 ($VI$), 7,174\,K  and 0.843 ($JHK$).
The dotted line shows the spectral energy distribution of the model
when only continuum processes are included.
}
\label{fig_fit_bb}
\end{figure}

\begin{figure}[htp!]
\epsfig{file=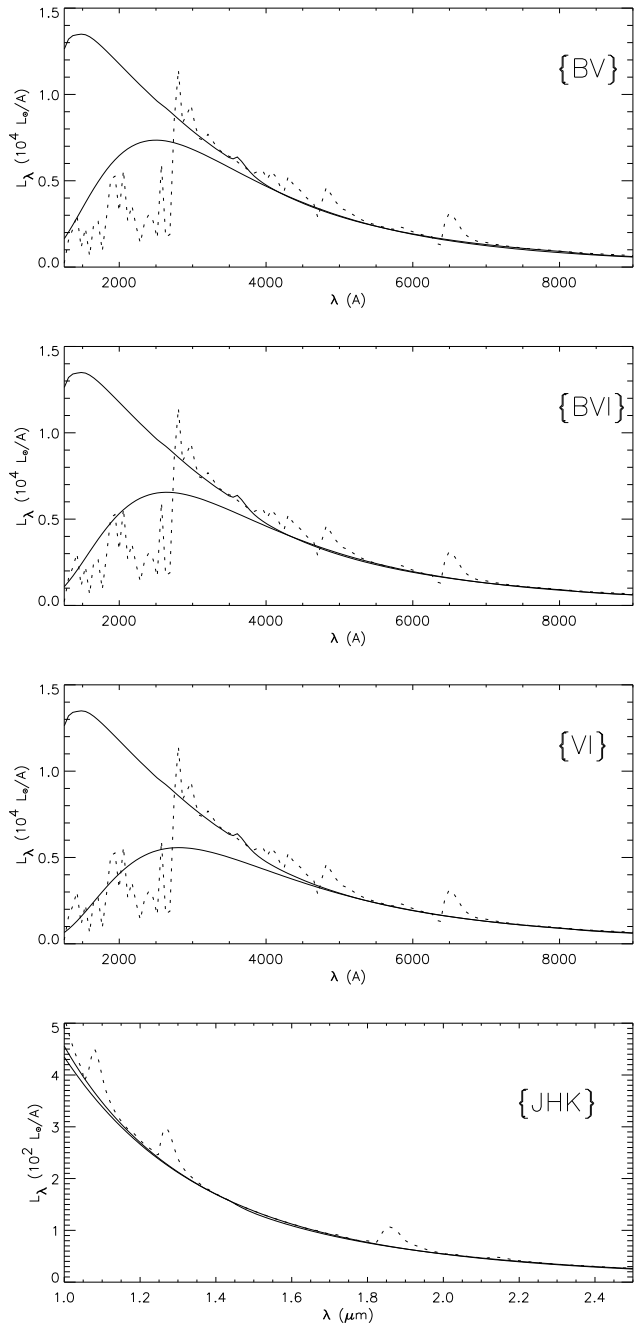, width=9cm}
\caption{
%
%
Same as previous figure but this time taking the continuum energy distribution.
The corresponding color temperatures and dilution factors are now:
11,599\,K and 0.486 ($BV$), 10,957\,K and 0.529 ($BVI$), 10,316\,K and 0.567 ($VI$) and
6,565\,K and 0.895 ($JHK$).
Note that both dilution factors and blackbody color temperatures are different
by about 10\% from the previous case where the synthetic spectrum contained
lines.
}
\label{fig_fit_bb_cont}
\end{figure}

\begin{figure}[htp!]
\epsfig{file=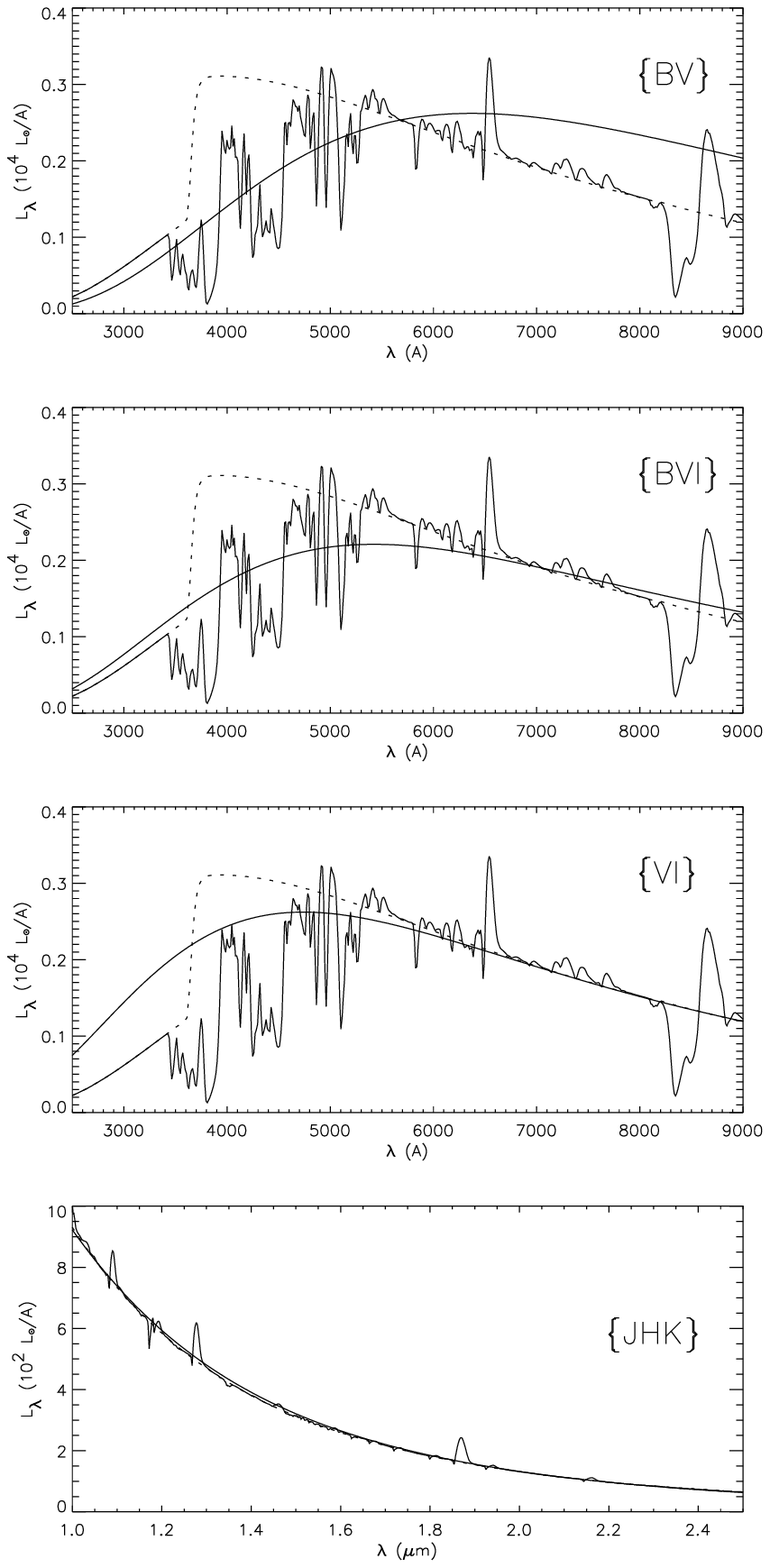, width=9cm}
\caption{
%
%
Blackbody fits in the four band passes to the model that reproduces well the
observations of SN1999em taken on the 22nd of November 1999 (see Dessart \& Hillier 2005c),
i.e. at ca. one month past core-collapse.
The corresponding color temperatures and dilution factors are now:
4,577\,K  and 1.471 ($BV$), 5,346\,K  and 0.914 ($BVI$), 6,180\,K and 0.7 ($VI$),
5,635\,K  and 0.790 ($JHK$).
Note how inadequate is the fit to the $BV$ set, resulting in a
correction factor greater than unity.
Conceptually, correction factors are {\it not} dilution factors since they
correct for a number of inadequacies of the EPM, presently the corrupting
effect of lines.
}
\label{fig_fit_bb_late}
\end{figure}

Let us now leave the discussion on correction factors and investigate the
adequacy of blackbody color temperatures obtained from the minimization
of $\epsilon$ given in Eq.~(\ref{eq9}).
In Fig.~\ref{fig_fit_bb}, we plot the blackbody SED (solid, featureless line)
that best fit the SED of a CMFGEN model ($T_{\rm eff}$ = 9,200\,K; solid line) 
over bandpass combinations $\{B,V\}$, $\{B,V,I\}$, $\{V,I\}$, and $\{J,H,K\}$.
Blackbody color temperatures (correction factors) derived for each set are, 
in the same order, 12,945\,K (0.443), 11,951\,K (0.494), 11,278\,K (0.526), and 
7,174\,K (0.843), thus above and below $T_{\rm eff}$ (Fig.~\ref{fig_ts_teff}).
The fit quality is good in all cases, although the blackbody  SED shows 
a sizable vertical shift, attributable to the presence of lines,
above the synthetic continuum SED (dotted line).
Indeed, $M_{\bar{\nu}}$, in Eq.~(\ref{eq9}), accounts indiscriminately for the {\it total
flux} in the corresponding bandpass, be it continuum, line, or a combination of both.
If, however, one uses, in Eq.~(\ref{eq9}), magnitudes computed on the synthetic 
continuum SED (Fig.~\ref{fig_fit_bb_cont}), the corresponding blackbody SED
then fits accurately the synthetic continuum SED.
The color temperatures are then lower and equal, in the same sequence, 
to 11,599\,K, 10,957\,K, 10,316\,K and 6,565\,K, with associated correction 
factors 5-10\% bigger.

The influence of lines on the optical SED strengthens as the outflow 
temperature further decreases.
A model with $T_{\rm eff} = $5,950\,K illustrates more dramatically 
this situation in Fig.~\ref{fig_fit_bb_late}: while the blackbody 
SED fits well the observations in the $\{J,H,K\}$ set, the agreement 
becomes poor for the sets $\{V,I\}$ and $\{B,V,I\}$, and unsatisfactory 
for the set $\{B,V\}$.
Here, the severe line-blanketing due to Fe{\,\sc ii} and Ti{\,\sc ii} (Paper I)
causes a major flux deficit around 4200\AA; the resulting blackbody color 
temperature, $T_{BV}$ = 4,545\,K, is too low to represent well the optical SED
(note the absence of a clearly defined continuum)
and correction factors, which adjust according to constraints from Eq.~(\ref{eq9}),
rise dramatically, as do the associated uncertainties.
Despite its weak influence on the effective temperature 
$T_{\rm eff}$ of the outflow, an enhanced Ti abundance could thus reduce 
the B-band flux and the inferred $T_{BV}$, mimicking a distance dilution.
Retaining observations sampling the early evolution of type II SN,
i.e. prior to the appearance of metal lines in the optical spectral range,
will therefore reduce the uncertainty on distances inferred with the EPM.

Within these limitations, the EPM should however deliver a distance to the object
with an accuracy set by the level of agreement between the synthetic and the observed 
fluxes (or magnitudes), since Eq.~(\ref{eq9}) provides the necessary correction to apply to 
a blackbody SED to retrieve, within numerical accuracy, the synthetic magnitudes for
any bandpass combination.
Overall, the above argument suggests that the EPM will be more accurate when combined
with a detailed {\it multi-wavelength} modeling of, rather than a degenerate 
2- or 3-point magnitude fit to the observed SED, thereby lifting possible
degeneracies. A further advantage of multi-wavelength modeling is that
the influence of lines can be more accurately gauged and hence allowed for.


\section{Dilution factors}
\label{Sec_dil_fac}

We now investigate the elements that contribute to the flux dilution caused
by the dominance of electron-scattering opacity.
In Sect.~\ref{Sec_dil_fac_ori}, we discuss its origin in more details.
A key quantity that controls the level of dilution is 
the spatial shift between the thermalization radius ($R_{\rm th}$) and 
the photospheric radius ($R_{\rm phot}$). 
We show in Sect.~\ref{Sec_dil_fac_phys} how these different radii 
vary with wavelength, and provide an additional discussion on the effect 
of line-blanketing.
In Sect.~\ref{Sec_dil_fac_dep}, we extend the discussion to a large set
of CMFGEN models, drawing statistical properties of $R_{\rm th}/R_{\rm phot}$
over a wide parameter space, i.e. corresponding to different density exponents,
outflow ionization, outflow velocity etc...
 

\subsection{Dilution factors and their origin}
\label{Sec_dil_fac_ori}

The origin of dilution factors has been discussed several times
in the literature (E96; Hershkowitz \& Wagoner 1987; 
Chilikuri \& Wagoner 1988; Schmutz et al. 1990). 
In order to clarify some concepts we repeat some of this discussion.
First, consider a plane parallel scattering atmosphere, and define $\lambda$ 
as the ratio of absorptive opacity $\kappa_{\rm abs}$ to the total opacity 
$\kappa_{\rm tot}$ (i.e., $\lambda = \kappa_{\rm abs}/\kappa_{\rm tot}$). 
If we assume that $\lambda$ is constant with depth, and that the Planck 
function varies linearly with $\tau$ (i.e, $B(\tau) = a + b \tau$) it can be 
shown that the mean intensity $J$ is given (approximately) by
  \begin{equation}
     J(\tau) = a + b \tau  + { {b - \sqrt{3} a} \over {\sqrt{3} +
     \sqrt{3\lambda}} }  \exp[ -\sqrt{3\lambda} \,\tau] 
  \end{equation}
(e.g., Mihalas 1978). We define the thermalization
optical depth, $\Lambda$, by 
\footnote{Mihalas defines the thermalization depth with unity instead of
0.67, while Schmutz et al. (1990) use 0.58 $(=1/\sqrt{3})$. We adopted
2/3 because we use that optical depth to define the photospheric radius.
These slightly different definitions are of no practical consequence.}
  \begin{equation}
    \Lambda = 0.67/\sqrt{\lambda}
  \end{equation}
Using this definition, and the previous expression it is apparent, 
for arbitrary $a$ and $b$, that $J$ will depart significantly from $B$ when the 
optical depth is less than the thermalization depth, i.e. $\tau < \Lambda$.

It can be further shown that the astrophysical flux F is given,
approximately, by
  \begin{equation}
        F = { 4 \over \sqrt{3} } {  \sqrt{\lambda} \over 1 + \sqrt{\lambda}}
            \,B(\tau= 1/\sqrt{3\lambda})
  \end{equation}
%
%
Thus the observed spectrum is that of a blackbody at $\tau= 0.87\Lambda$
but diluted by a factor 
  \begin{equation}
     {4 \over \sqrt{3} } { \sqrt{\lambda} \over {1 + \sqrt{\lambda}} }
  \end{equation}
For a strongly scattering atmosphere (i.e, $\lambda << 1$), the dilution 
($=\xi^2$) is approximately 2$\sqrt{\lambda}$.

Unfortunately, the situation with a SN is much more complicated.
First, $\lambda$ is depth dependent, since 
the scattering opacity (due to free electrons) scales linearly with density
while the absorptive opacity, due to bound-free and free-free processes,
scales typically with the square of the density.
However, we can define an approximate thermalization optical depth scale
by
  \begin{equation}
    \tau_{\rm th}(r)  =  \int_r^\infty \sqrt{\lambda} \, \kappa_{\rm tot} \,dr .
    \label{eq_tau_therm}
  \end{equation}
The thermalization layer is located on this optical depth scale at a value of 2/3.
Since $\lambda$ varies with wavelength, $\tau_{\rm th}$ also does. Consequently the 
observed spectrum is not that of a pure blackbody.

Second, SN atmospheres have a finite extent. Thus, the photospheric radius 
$R_{\rm phot}$ (defined, using the plane-parallel results, as occurring at 
$\tau=2/3$) is larger than the radius of the SN at the thermalization 
depth ($R_{\rm th}$). 
This introduces an additional dilution of the radiation --- roughly as 
$(R_{\rm th}/R_{\rm phot})^2$.

Third, the finite extent of the envelope also introduces an additional correction due
to emission from the overlying layers.

Given these complications it makes sense to write, as done in Eq.~(3),
\beq
   f^{\rm obs}_\nu = \xi^2 \, \theta^2 \, 10^{-0.4 A_\nu} \, \pi B_\nu(T_c)
\eeq
where $\xi$ represents a factor that corrects for the measured blackbody flux 
to give the observed flux. 
Note that with this approach, the precise definition of photospheric radius 
is unimportant --- what is important is that we compute the $\xi$ for 
whatever definition we use.
As discussed in Sect.~\ref{Sec_corr_fac}, the correction factors entering the
EPM are not simply accounting for the effects just discussed, but also for
the significant impact that line-blanketing has on the SED. 

In the next section we illustrate how the thermalization and photosphere radii 
vary with wavelength. In terms of the thermalization radius we find that the 
total dilution ($\xi^2$) scales as 
$$
(R_{\rm th}/R_{\rm phot})^{n+1}
$$
where $n$ is the density exponent in the SN, and we have assumed that the ratio of 
absorption to scattering opacity is constant. Thus, a small change in thermalization
radius can produce a substantial change in the dilution factor.

\subsection{Wavelength variation of $R_{\rm th}$ and $R_{\rm phot}$ in realistic SN models}
\label{Sec_dil_fac_phys}

\begin{figure}[htp!]
\epsfig{file=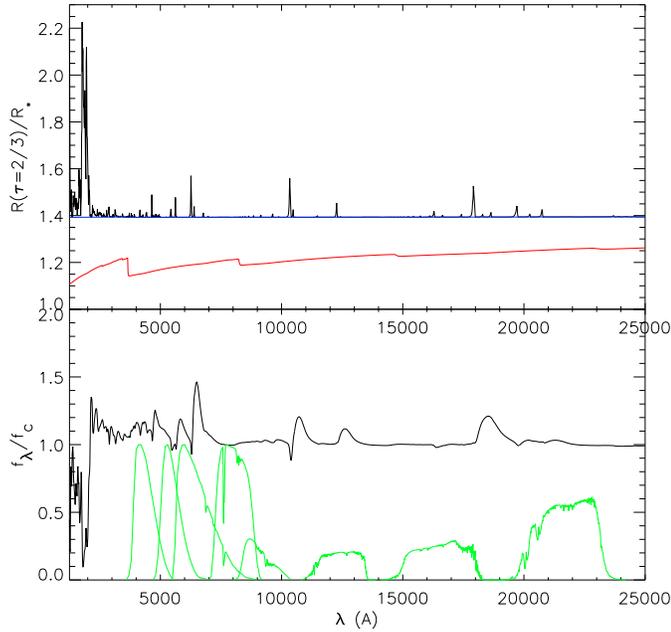, width=9cm}
\caption{
{\it Top panel}:
wavelength variation of the radius at optical-depth two-third,
including from the bottom to the top curves, only bound-free and
free-free processes (red), bound-free plus free-free plus electron-scattering
contributions (blue), and all continuum and line processes together (black).
The effective temperature of the model is $T_{\rm eff} =$13,610\,K.
{\it Bottom panel}:
corresponding wavelength variation of the rectified flux.
Also overplotted are the transmission curves of the $B$, $V$, $I$, $J$ $H$,
and $K$ bands from Bessell (1990, 1988) and H01. The rectified flux was
obtained by dividing the model spectrum by the model 
continuum spectrum.
}
\label{fig_dil_early}
\end{figure}

   We have selected three models covering the early-, intermediate- and
late-stage evolution of a type II SN during its photospheric phase.
In the top panel of Fig.~\ref{fig_dil_early}, we plot the wavelength variation
of the radius of optical depth two-third for a model with $T_{\rm eff}=$13,610\,K,
normalized to the base radius $R_0$ (also noted $R_{\ast}$) of the model grid.
Three curves are illustrated:

\begin{enumerate}
\item
 $R(\tau=2/3)$ when all processes, such as bound-bound, bound-free,
                 free-free, and electron-scattering are included
                 (black curve).
\item
 $R(\tau=2/3)$ when all processes, except bound-bound transitions, are
                 included (blue). This defines the classic photosphere.
                 In most regions, it is set by the electron scattering
                 opacity.
\item
 $R(\tau_{\rm th}=2/3)$ where $\tau_{\rm th}$ is computed using
      Eqn.~(\ref{eq_tau_therm}) and we assume that the
      scattering opacity is due to electrons, while the absorptive opacity
      is due to bound-free and free-free processes only (red).
\end{enumerate}

Several elements are apparent in this plot.
First, the curves depart from the horizontal by various degrees, i.e.
the variation with wavelength depends on the processes included.
The thermalization radius, although generally 
increasing with wavelength,  shows a sawtooth behavior.
The main opacity process that links photons to the thermal pool is the
photo-ionization of hydrogen, the sudden jumps in thermalization radius
at 3649\AA\, (Balmer jump) and 8211\AA\, (Paschen jump) resulting from 
photo-ionization from the $n=2$ and $n=3$ levels of hydrogen.
The thermalization radius varies between 1.1 and 1.25\,$R_0$ where $R_0$ is the
base radius of the model.
%
%
While this represents a modest radius variation, the corresponding
electron-temperature decreases from 25,000\,K to 17,000\,K between these two
heights so that the associated Planckian photon distribution at 1000\AA\, and
25000\AA\, at the thermalization depth are in fact substantially different.
The second curve up is remarkably wavelength independent, fixed at 1.4\,$R_0$.
It corresponds to the photosphere, for which one usually assumes
only continuum processes.

We show in the bottom plot the rectified flux over the same wavelength
range as that of the top panel.
Note how the flux minima and maxima are ``echoed" in the top panel with large increases
in radius. This is particularly true in the UV where metal line-blanketing
around 2000\AA\, is strong.
Thus, strictly speaking, the spatial size of the photosphere
is very dependent on wavelength, but this variation has different
magnitudes depending on the selected opacity sources and the wavelength region
considered.

\begin{figure}[htp!]
\epsfig{file=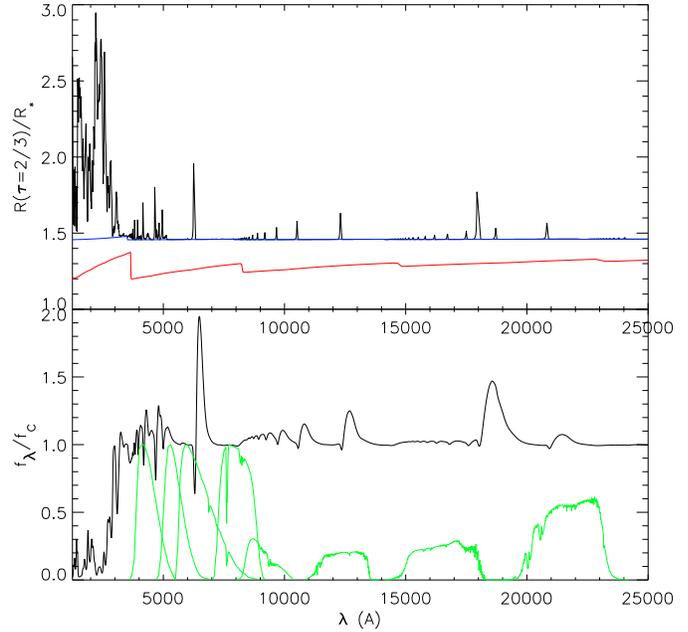, width=9cm}
\caption{
Same as Fig.~\ref{fig_dil_early} for a cooler model (intermediate-stage in
the photospheric-phase evolution), with $T_{\rm eff} =$8,035\,K.
The maximum ordinate in the top panel is 3, rather than 2.3 as 
in Fig.~\ref{fig_dil_early}.
}
\label{fig_dil_int}
\end{figure}

\begin{figure}[htp!]
\epsfig{file=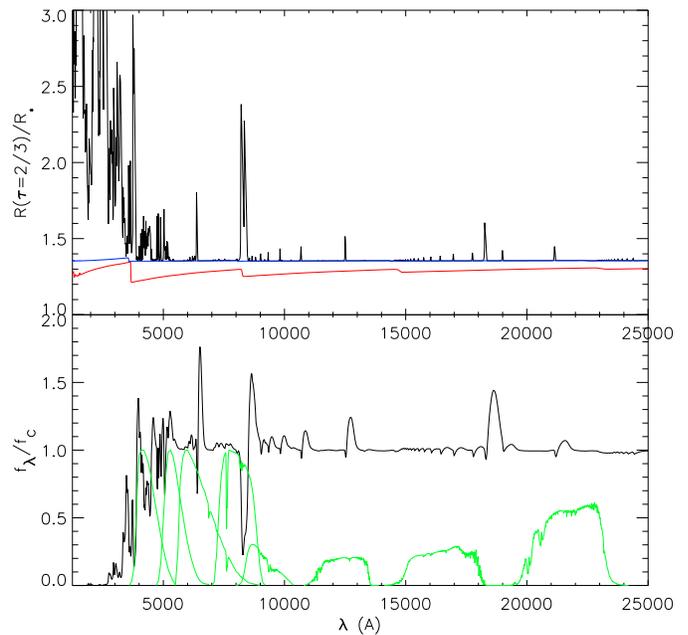, width=9cm}
\caption{
Same as Fig.~\ref{fig_dil_early} for a much cooler model (late-stage in the
photospheric evolution), with $T_{\rm eff} =$6,460\,K.
The maximum ordinate in the top panel is 3, rather than 2.3 as 
in Fig.~\ref{fig_dil_early}.
}
\label{fig_dil_late}
\end{figure}

  In Figs.~\ref{fig_dil_int} and \ref{fig_dil_late}, we show a similar set
of plots but now for models of decreasing effective temperature (outflow 
ionization), corresponding to a later stage of SN evolution where hydrogen
recombines at or above the photosphere.
One key feature shown in these plots is that the thermalization radius 
is now much closer to the photospheric radius: the reduced density of free-electrons favors the
contribution of bound-free and free-free processes to the total optical depth at the 
photosphere.
%
%

A second effect clearly appears in these latter two figures, namely
the presence of lines, first predominantly in the UV
(Fig.~\ref{fig_dil_int}), and then also in the optical range (Fig.~\ref{fig_dil_late}).
The effects of metal lines in the UV increases the photosphere radius from
ca. 1.3\,$R_0$ up to ca. 2-4\,$R_0$. In the optical range, the $B$ and
to a lesser extent the $V$ bands become more and more contaminated by lines so that
the effective photospheric radius is 10-15\% bigger than predicted when accounting
only for continuum opacity.
So, the first effect of lines at later times is to spread the photosphere
radius over a range of values, making it non-unique and therefore
less adequate for the application of the Baade-Wesselink method.
But as the outflow cools down, lines become ever stronger and
it becomes more and more dubious to approximate the SED with a
blackbody distribution of whatever temperature (EPM).
In the I band, which at early times was completely devoid
of lines, the Ca{\,\sc ii} line around 8800\AA, causes a huge variation
in the (true) photosphere radius of cooler models.
It is again tempting to advise not to use the I band at later times for the EPM.

\begin{figure*}[htp!]
\epsfig{file=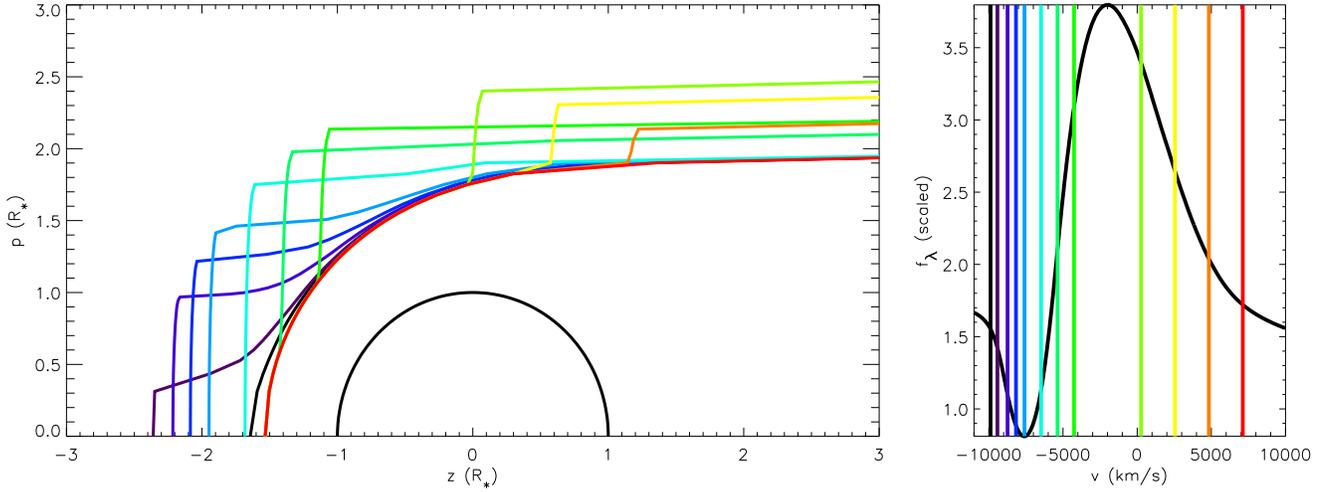, width=18cm}
\caption{
{\it Left}: Optical-depth-two-third contours for H$\alpha$, shown
in the $(z,p)$ plane, for an observer located at $p=0$ and $z=-\infty$.
Each contour corresponds to a given Doppler-shift velocity from the
rest wavelength of $\lambda_0 = 6562.79$\AA.
{\it Right}: Emergent H$\alpha$ profile shown in velocity space, i.e.
scaled wavelength displacements from the rest-wavelength
$(\lambda/\lambda_0-1)c$.
To assist with the visualization, each contour on the left corresponds
to a vertical line of the same color in the right panel that gives off the
selected velocity away from line-center. The red curve
is for the continuum which, for practical purposes, is constant across the line [color].
}
\label{fig_pz}
\end{figure*}

Let us illustrate further the strong impact lines can have on modifying the photospheric
radius.
We show in Fig.~\ref{fig_pz} contours of total optical depth two-third
in the $(z,p)$ plane (in units of $R_0$)
at selected positions within the H$\alpha$ line profile, placing the observer
at infinity to the left ($p=0, z =  -\infty$).
The black half-circle corresponds to the base of the grid at $R_0$.
Here, the optical depth includes all processes, but over the small wavelength
region considered, only the contribution from the line optical depth varies.
Each contour corresponds to a given wavelength displacement from line center,
shown as a vertical line of the same color in the right panel.
Throughout the profile, the minimum optical depth is fixed by continuum
processes, which, as said before, are essentially constant over the line profile.
This corresponds to the curve that is closest to the inner half-circle,
shown in red.
Any departure from that curve is due to line-optical depth effects.

In outflows characterized by a Hubble velocity, the iso-velocity contours
(i.e, contours on which the projected line of sight velocity is constant)
coincide with $z$=constant (i.e., planes perpendicular to our line of sight).
Thus, if the line optical depth was only dependent on velocity, the
iso-velocity and iso-$\tau$ contours would overlap.
However, the line optical depth decreases with the density, which drops
as a high power of the radius in SN outflows.
Thus, an iso-velocity (or constant-z) surface cuts iso-density contours
for increasing values of $p$.
Beyond a given maximum impact parameter, the line optical depth becomes
negligible and the total optical depth takes its minimum value.
For sufficiently high values of $p$, there is no surface of total
optical depth unity.
The last such surface defines the limb of the star.

At blue-wavelengths from line center, the iso-$\tau$ contour is strongly
peaked along $p=0$.
At wavelengths closer to line center, the iso-$\tau$ surfaces cover a bigger
range of impact parameters, going up to a value of 2.3\,$R_0$ which corresponds
to the wavelength for maximum flux (line center region).
This means that the radiating surface at 6562.79\AA\, is 70\% bigger
than that just redward of H$\alpha$ where no lines are present.
We thus see that the radiating surface has a wavelength-variable radius
once lines are accounted for, invalidating further the assumption
of constant photospheric radius and blackbody energy distribution.
These facts support the idea that the Baade-Wesselink method will
work best during the early-stage of the photospheric phase evolution
of type II SN, when they show a relatively featureless spectrum and a more
uniform dilution from electron scattering.

\subsection{Statistical properties of $R_{\rm th}/R_{\rm phot}$ in realistic SN models}
\label{Sec_dil_fac_dep}


In the previous section, using a set of three CMFGEN models of distinct
outflow ionization, we have discussed the wavelength variation of the location
of selected radii, corresponding to thermalization, continuum and total optical depth
two-third.
If lines are ignored, the photospheric radius is essentially unchanged throughout
the spectrum.
If lines are accounted for, as they should be, the resulting ``true'' photospheric 
radius can vary widely over small wavelength ranges.
For the following discussion 
we make the important simplification that the photosphere is only set 
by continuum opacity processes.
This will be inaccurate at intermediate- and late-times but reasonable
during the early-phase of photospheric evolution of type II SN (i.e. 
until metal-lines appear in the optical range).
Below we investigate how the quantity $R_{\rm th}/R_{\rm phot}$ varies over
a wide range of outflow properties, controlled, e.g., by outflow ionization,
the density distribution, and the spatial scale.
Using the modest variation of $R_{\rm th}$ with wavelength, we simplify the
discussion by building optical and near-IR averages (i.e. over the 
sets $\{B,V,I\}$ and $\{J,H,K\}$) for $R_{\rm th}/R_{\rm phot}$.

\subsubsection{Dependence on $T_{\rm eff}$ and $\rho_{\rm phot}$}
\label{Sec_dil_fac_dep_tr}

\begin{figure}[htp!]
\epsfig{file=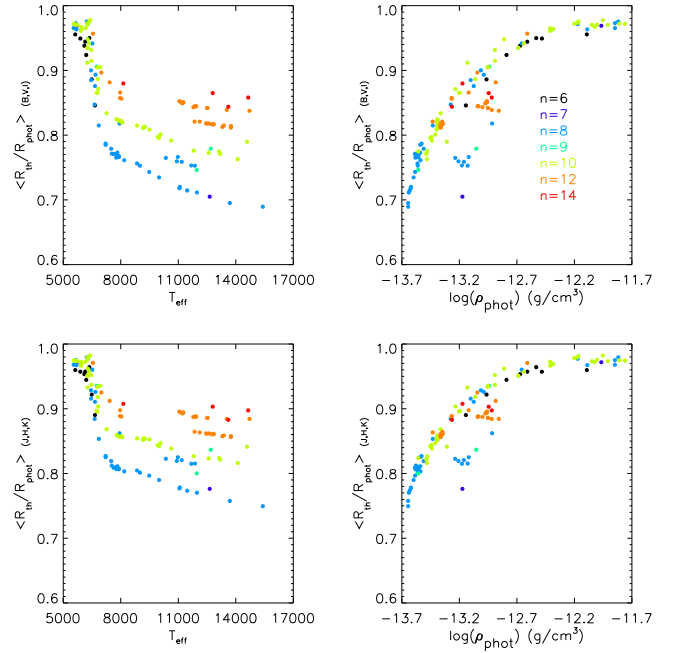, width=9cm}
\caption{
Plot of $R_{\rm th}/R_{\rm phot}$ averaged over the two sets $\{B,V,I\}$ and $\{J,H,K\}$
versus effective temperature (left) and photospheric density (right) for
142 CMFGEN models covering a range of outflow ionization pertinent for
modeling the
photospheric evolution of type II SN.
To partially explain the scatter of points, we use a color-coding that groups
models
that have the same density exponent (see
Sect.~\ref{Sec_dil_fac_dep_tr}-\ref{Sec_dil_fac_dep_other}).
Note that a number of models which do differ in spectroscopic appearance have
comparable $R_{\rm th}/R_{\rm phot}$, leading to numerous overlap of data points.
}
\label{fig_dil_quad}
\end{figure}

In Fig.~\ref{fig_dil_quad}, we first illustrate the variation
of $R_{\rm th}/R_{\rm phot}$ with $T_{\rm eff}$ and $\rho_{\rm phot}$,
using averages over the two sets $\{B,V,I\}$ and $\{J,H,K\}$.
Let us focus on the $\{B,V,I\}$ set and discuss the results for the
$n=10$ models shown as green dots (we leave the discussion for the
dependence on $n$ until the next section).

For this sub-set, there is a very modest scatter of $R_{\rm th}/R_{\rm phot}$
in the temperature range 8,000 to 17,000\,K, by $\pm$0.03 around ca. 0.68.
At the low temperature end, this ratio rises towards unity.
This is easily understood by recalling the previous section.
Above approximately 8,000\,K, the outflow temperature and ionization is such
that most electrons are unbound and thus contribute a significant
and more or less fixed fraction to the total continuum optical-depth
at the photosphere.
At approximately 8,000\,K, hydrogen suddenly recombines in the photosphere region and the
density of free-electrons drops. For lower photospheric temperatures,
the electron-scattering contribution to the total continuum optical-depth
at the photosphere decreases in favor of bound-free and free-free opacity sources,
so that the corresponding dilution factors increase towards unity.
This limiting case is reached when electron-scattering no longer
contributes to the optical depth at the photosphere, i.e. when
no free-electrons are present.
In this case, the SED at the photosphere suffers no dilution; indeed, 
it is a fundamental property that $R_{\rm th}/R_{\rm phot}$ be less than unity. 

When $R_{\rm th}/R_{\rm phot}$ is plotted as a function of photospheric density, the
models with $n=10$ (green-dots) also show a tight correlation,
falling on a positive-gradient line with increasing density.
As the density increases, $R_{\rm th}/R_{\rm phot}$ increases towards unity.
This occurs since a higher and higher photospheric density is required to allow 
bound-free and free-free
processes to compensate for the reduced contribution of electron-scattering
at the photosphere (due to recombination of hydrogen).
Moving to the lower panel and the corresponding illustrations for the set
$\{J,H,K\}$, we find that $R_{\rm th}/R_{\rm phot}$ shows a very similar behavior,
with values being merely shifted upwards by up to ca. 0.1.

The tight correlation of $R_{\rm th}/R_{\rm phot}$ with photospheric density
and effective temperature is not clearly interpreted since these quantities
may not be completely independent.
To disentangle potential combined effects, we show in Fig.~\ref{fig_dil_cont_n10}
a contour plot for $R_{\rm th}/R_{\rm phot}$ averaged over the $\{B,V,I\}$ set,
and limited still to models with $n=10$ (green dots in Fig.~\ref{fig_dil_quad};
note also that similar findings would be obtained for the near-IR set).
As before, we find that in the fully ionized models
there is a pronounced degeneracy of $R_{\rm th}/R_{\rm phot}$
with temperature, as revealed by near-horizontal contours.
Points with $R_{\rm th}/R_{\rm phot} \la 1$ occur in the
parameter space where the $T_{\rm eff}$ is low and $\rho_{\rm phot}$ is high,
i.e. higher $R_{\rm th}/R_{\rm phot}$ are found as one moves from the lower right
corner (low $\rho_{\rm phot}$ and high $T_{\rm eff}$) to the upper left corner
(high $\rho_{\rm phot}$ and low $T_{\rm eff}$).
As discussed above, as one moves from fully ionized models
to cooler ones where hydrogen recombines in the outflow, the electron-scattering
optical depth at the photosphere is significantly lower due to the reduced
density of free-electrons.
For reduced photospheric temperatures, higher photospheric densities are required
to compensate for the reduced contribution of electron-scattering to the
continuum optical depth, explaining the rise of $R_{\rm th}/R_{\rm phot}$
from the bottom-right corner to the upper-left corner of Fig.~\ref{fig_dil_cont_n10}.



\subsubsection{Dependence on secondary parameters}
\label{Sec_dil_fac_dep_other}

Let us now broaden the discussion of the previous section by investigating
the dependence of $R_{\rm th}/R_{\rm phot}$ on model parameters at large, and not just
the density and temperature at the photosphere.
First, let us go back to Fig.~\ref{fig_dil_quad} and consider all points.
The scatter at a given temperature is then about 3-4 times as large as in the case
where a single color is considered.
When the color-coding is employed, models with a given density exponent are
grouped together and the resulting scatter of points is quite small.
The effect is easily understood if one considers the ideas discussed in
Sect.~4.5 of Paper I.
Thermalization processes are density-square dependent, while electron-scattering
is linear in the density.
Lowering (increasing) $n$ makes the density in the outer layers larger (smaller) 
and thus leads to a lower (larger) density at the photosphere: the opacity 
due to thermalization processes is lowered (increased) and $R_{\rm th}/R_{\rm phot}$ 
is reduced (enhanced).
In general, a lower $n$ favors the contribution of electron-scattering
in fully ionized models and therefore results in low values of $R_{\rm th}/R_{\rm phot}$.
In cool models, the density exponent has a less pronounced importance since the outflow
optical depth is essentially set by the optical depth of the hydrogen-recombination
front. This typically extends over a spatially confined region, while the outer
regions contribute a negligible electron-scattering optical depth due to the
low density of free-electrons.

In the right-hand side of Fig.~\ref{fig_dil_quad}, a number of models fall
below the curved line followed by most models.
These correspond to models that have a different
scale, i.e. smaller luminosity and radius but similar outflow ionization
(same T$_{\rm eff}$).
For example, setting the base radius at $R_0$ and 10\,$R_0$, the base density
at $\rho_0$ and $\rho_0 / 10$ and the luminosity as $L_0$ and 100\,$L_0$
gives two models with identical photospheric velocity and effective temperature
and identical electron-scattering optical depth at the grid base.
However, the model with a larger radius has $R_{\rm th}/R_{\rm phot}$ smaller by ca. 10\%
(enhanced flux dilution) than that of the model with the smaller radius.
This is explained in the same terms as above for the models with different $n$.
For more compact objects, the reduced scale of the outflow reduces the contribution
from electron scattering (which varies linearly with the spatial scale) and leads to
a higher density at the photosphere.
Also, we find no noticeable metallicity effect on the resulting $R_{\rm th} / R_{\rm phot}$.
This is easily understood --- in type II SN the thermalization processes stem from 
the photoionization of hydrogen atoms which is unrelated to the metal opacity.
Moreover, modulating the model expansion velocity does not change the behavior
of $R_{\rm th} / R_{\rm phot}$.

\begin{figure}[htp!]
\epsfig{file=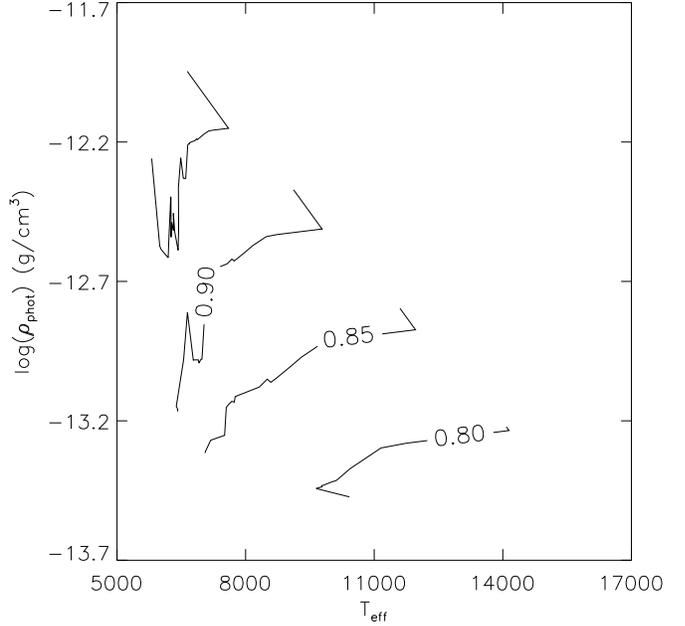, width=9cm}
\caption{
Contour plot of $<R_{\rm th}/R_{\rm phot}>_{\{B,V,I\}}$  in the $T_{\rm eff}-\rho_{\rm phot}$ plane
for models with $n=10$, shown as green dots in Fig.~\ref{fig_dil_quad}.
Iso-contours are this time shown for values of 0.80, 0.85, 0.90, and 0.95.
}
\label{fig_dil_cont_n10}
\end{figure}

\begin{figure}[htp!]
\epsfig{file=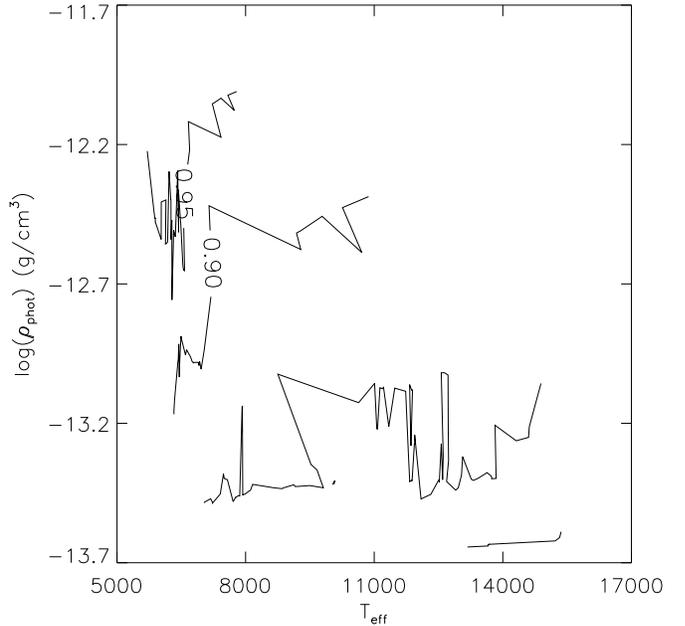, width=9cm}
\caption{
Contour plot of $<R_{\rm th}/R_{\rm phot}>_{\{B,V,I\}}$  in the $T_{\rm eff}-\rho_{\rm phot}$ plane.
For visibility purposes, we only show iso-contours for $\xi$-values of 0.7, 0.8, 0.9, and 0.95.
}
\label{fig_dil_cont}
\end{figure}

Finally, in Fig.~\ref{fig_dil_cont}, we duplicate Fig.~\ref{fig_dil_cont_n10} by including all
the models shown in Fig.~\ref{fig_dil_quad}.
The scatter is significantly enhanced, although the general trend follows the same
rule as observed with models having $n=10$ only.
So, while small variations from model to model can result from changes,
mostly, on the density exponent and the spatial scale, the general behavior
of $R_{\rm th}/R_{\rm phot}$ is always towards higher, i.e. closer-to-unity, values for
cooler temperatures and concomitant higher photospheric densities.


\section{Inference of the photospheric velocity}
\label{Sec_vphot}

To infer the distance to a type II SN, the EPM requires the evaluation
of the photospheric velocity at each observation date.
The standard procedure is to adopt the velocity at maximum absorption $v_{\rm abs}$ in
P-Cygni profile troughs, favoring the more optically-thin lines (L02).
This measurement has an inherent uncertainty, particularly at early times when profile
troughs show a weak curvature; besides, it is unclear how suitable this location is
since line profile formation and shape result from complex line and
continuum optical-depth effects (Sect. 5, Paper I).
In this section, we address the reliability of the above procedure by comparing
synthetic line-profile velocity measurements with the corresponding model photospheric velocity.

\begin{figure}[htp!]
\epsfig{file=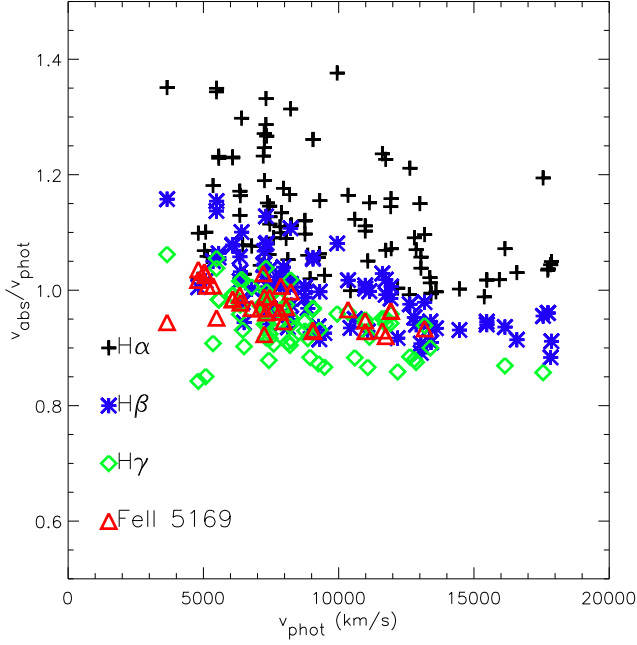, width=9cm}
\caption{Plot based on measurements performed on synthetic spectra
of $v_{\rm abs}/v_{\rm phot}$ as a function of $v_{\rm phot}$ for
H$\alpha$, H$\beta$, H$\gamma$, and the Fe{\,\sc ii} 5169\AA\, feature.
$v_{\rm abs}$ corresponds to the velocity at maximum absorption
in the corresponding P-Cygni line profile. $v_{\rm phot}$ is the
model velocity at which the total continuum optical depth is 2/3.
Note that there are fewer data points for the Fe{\,\sc ii} feature
since it is only observable in relatively cooler outflows, which
statistically possess a lower expansion velocity [color].
}
\label{fig_vel}
\end{figure}
\begin{figure}[htp!]
\epsfig{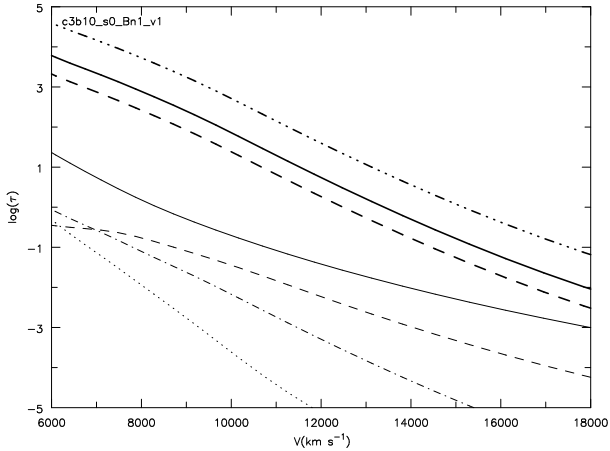}
\caption{
Plot of line (thick lines; dashed-dotted: H$\alpha$; solid: H$\beta$; dashed:
H$\gamma$) and continuum (thin line; solid: electron-scattering; dashed:
continuum at 1000\AA; dashed-dotted: continuum at 3000\AA; dotted:
continuum at 5000\AA) optical depth as a function of velocity for the model
discussed in Sect.~\ref{Sec_vphot}.
The photosphere is located at a velocity of 8750\,km\,s$^{-1}$,
at which the optical depth in hydrogen Balmer lines is orders of magnitude
larger than that in the continuum.
}
\label{fig_tau}
\end{figure}

In Fig.~\ref{fig_vel}, using a sample of ca. one hundred photospheric-phase type II SN models,
we show the velocity measurement at maximum absorption in synthetic line profiles
of H$\alpha$, H$\beta$, H$\gamma$, and Fe{\,\sc ii} 5169\AA, normalized to and as a function
of the model photospheric velocity $v_{\rm phot}$ - statistically, models that have a higher
$v_{\rm phot}$ also have a higher $T_{\rm eff}$ and correspond to times closer to the explosion date.
If the velocity location at maximum absorption in the synthetic line profile corresponded
accurately with the model photospheric velocity, all data points would gather along the
$v_{\rm abs}/v_{\rm phot}=1$ line.
Instead, we see a moderate scatter, between 0.8 and 1.4, enforced by the strong confinement
within the photospheric layer of line and continuum formation (Sect. 5, Paper I).
Nonetheless, this scatter, a function of the synthetic line chosen, is large from the
perspective of the EPM (Sect.~\ref{Sec_pres_epm}), with points both {\it above} 
and {\it below} this unity-line, implying a potential {\it overestimate} or {\it underestimate} 
of the photospheric velocity.
For H$\alpha$, most data points lie above the unity-line and thus adopting $v_{\rm abs}$
for that line generally overestimates $v_{\rm phot}$, by up to 40\%, but only marginally
for high-$v_{\rm phot}$/high-$T_{\rm eff}$  models.
For H$\beta$ and H$\gamma$, data points show a similar scatter as H$\alpha$,
but shifted down by ca. 20\%, matching $v_{\rm phot}$ by $\pm 15$\%.
But for the Fe{\,\sc ii} 5169\AA\, line, the velocity measurement $v_{\rm abs}$
matches $v_{\rm phot}$ to within 5-10\%.

This suggests that, when applying the EPM or the SEAM, setting $v_{\rm phot}$
equal to the velocity at maximum absorption in the Fe{\sc ii}\,5169\AA\, line
achieves high accuracy, as argued, e.g., by Schmutz et al. (1990).
Unfortunately, this line is only present in synthetic spectra of models with relatively
low $T_{\rm eff}$, corresponding to late times after explosion, when the use of
either the SEAM or the EPM is not optimal
(Sect.~\ref{Sec_pres_epm}-\ref{Sec_corr_fac}-\ref{Sec_dil_fac}).
At early times, however, $v_{\rm phot}$ can be inferred from hydrogen Balmer line 
measurements, but with a larger associated uncertainty that is difficult to estimate 
by eye-ball (Fig.~\ref{fig_vel}).
Thus, overall, the use of theoretical models like CMFGEN can contribute to reducing the
uncertainty on the inferred photospheric velocity by ca. 10-20\% compared to
methods based exclusively on direct measurement of observed line profiles,
a significant gain given the desired accuracy for subsequent distance determinations.

Despite the relatively good agreement obtained above 
(at the few tens of percent level), $v_{\rm abs}$ can, in some models, 
be lower than $v_{\rm phot}$.
This is contrary to intuition. Since these lines have to be optically thicker 
than the continuum to leave an imprint on the emergent spectrum, one expects the 
line-flux minimum to occur exterior to, and thus at higher velocities than that at 
the photosphere.  
The possible underestimate of $v_{\rm phot}$ with the use of $v_{\rm abs}$
can be inferred from the work of Branch (1977), and Jeffery \& Branch (1990), 
but no explanation is given.

To understand the behavior of the velocity minima,
consider the example of the intermediate-stage model of Paper I (Sect. 3.2).
In Fig.~\ref{fig_tau}, we see that the total optical depth of hydrogen Balmer lines
(thick solid curves) is orders of magnitude larger than that in the continuum
(solid or broken, thin curves).
Yet, measurements on the synthetic spectrum (Fig. 3, Paper I)  yield  $v_{\rm abs}$/$v_{\rm phot}$ of
1.10, 0.99, and 0.93 for H$\alpha$, H$\beta$, H$\gamma$,
confirming the typical trend discussed in Fig.~\ref{fig_vel}.
We shall see below that the underestimate/overestimate of $v_{\rm phot}$ is tied to
the way the emergent line intensity varies with depth and impact parameter $p$.



\begin{figure}[htp!]
\epsfig{file=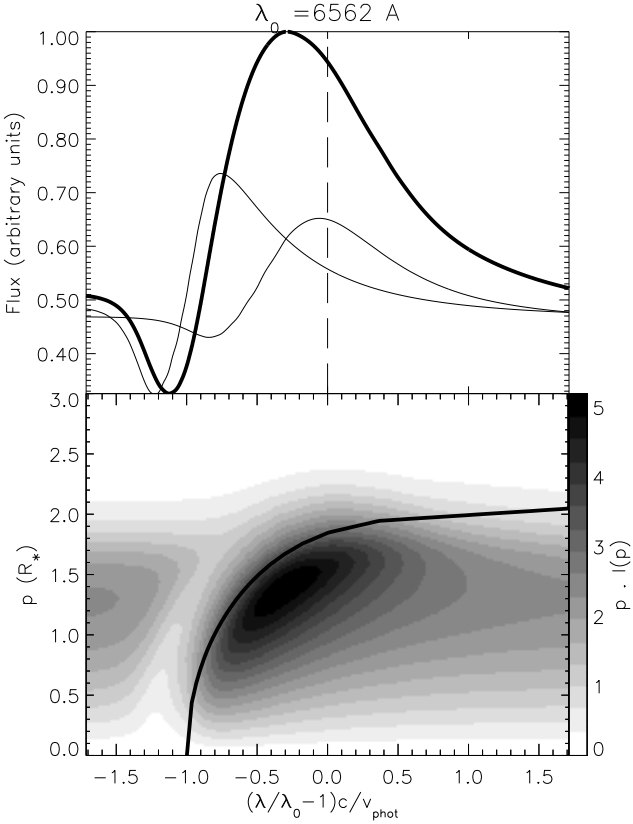, width=8cm}
\caption{
{\it Bottom}: grayscale image of the quantity $p \cdot I(p)$ as a
function of $p$ and scaled wavelength $x=(\lambda/\lambda_0-1)c/v_{\rm phot}$,
where $p$ is the impact parameter (in units of the hydrostatic radius R$_{\ast}$)
and $I(p)$ the emergent specific intensity for a ray with $p$ (at $x$).
Here, $\lambda_0$ corresponds to the rest wavelength of H$\alpha$ and $c$ is the speed of light.
The photospheric velocity of this model is 8750\,km\,s$^{-1}$.
The overplotted thick black curve gives the line-of-sight velocity location where the
integrated continuum optical depth at 5500\AA, along $z$ and at a given $p$, equals 2/3.
{\it Top}: line profile flux, directly obtained by summing $p \cdot I(p)$ over the
range of $p$.
Overplotted are the curves $p \cdot I(p)$ for $p=0.05 R_{\ast}$ and $p = 1.8 R_{\ast}$
(thin lines; scaled arbitrarily and individually for clarity).
}
\label{fig_jnu2_halpha}
\end{figure}

Let us first focus on H$\alpha$.
In the bottom panel of Fig.~\ref{fig_jnu2_halpha}, we show the gray scale image of the flux-like quantity
$p \cdot I(p)$ as a function of scaled wavelength $x=(\lambda/\lambda_0 - 1)c/v_{\rm phot}$
and $p$. For clarity in this section, $I(p)$ is computed by accounting only for the 
bound-bound transition in question. Overplotted (thick line) is a curve showing the location of
the continuum photosphere, $(x_{\rm phot,p},p)$.
Along a given ray $p$, $p \cdot I(p)$ shows a minimum at $x=x_{{\rm min},p}$, with
$x_{{\rm min},p} < x_{{\rm phot},p}$.
This is to be expected because of the greater optical depth in the line than in the continuum,
and because of the lower H$\alpha$ source function (Paper I).
However, following the curve $(x_{\rm phot,p},p)$ as a guide, we find that the
location $x_{{\rm min},p}$ varies greatly with $p$. This has important implications
for the location of the absorption minimum in the observed profile.

In the top panel of Fig.~\ref{fig_jnu2_halpha}, beside the total emergent line profile (solid line),
we plot the quantity $p \cdot I(p)$ (scaled for clarity) for two selected impact parameters,
$p=0.05 R_{\ast}$ (impacting close to the SN center) and $p=1.8 R_{\ast}$ (the
limiting $p$ where both line and continuum become optically-thin for all depths along the ray).
For the inner ray, the minimum occurs at $|x| \sim 1.25$, thus at a velocity well above $v_{\rm phot}$;
but for the outer ray, it occurs at $|x| \sim 0.8$, thus for a velocity well below it.
When building the total profile, one combines contributions from rays with impact parameters
encompassing this range. The location of the minimum in the resultant profile will
thus depend on the relative weighting of different impact parameters, which will be transition
dependent.

The cause of the variation in the location of the absorption maximum
was partially explained in Sect.~\ref{Sec_dil_fac}:
iso-velocity curves are at constant $x$ (depth $z$ in the $(p,z)$ plane)
but the density varies as $1/r^n$.
Thus, at fixed $z$, the density drops fast for increasing $p$, at the same time
reducing the probability of line scattering and/or absorption.
In practice, the location of maximum absorption along a ray shifts to larger depths
($z$ closer to zero) for increasing $p$, showing overall the same curvature as seen for
the continuum photosphere (see also Fig.~\ref{fig_pz}).

\begin{figure}[htp!]
\epsfig{file=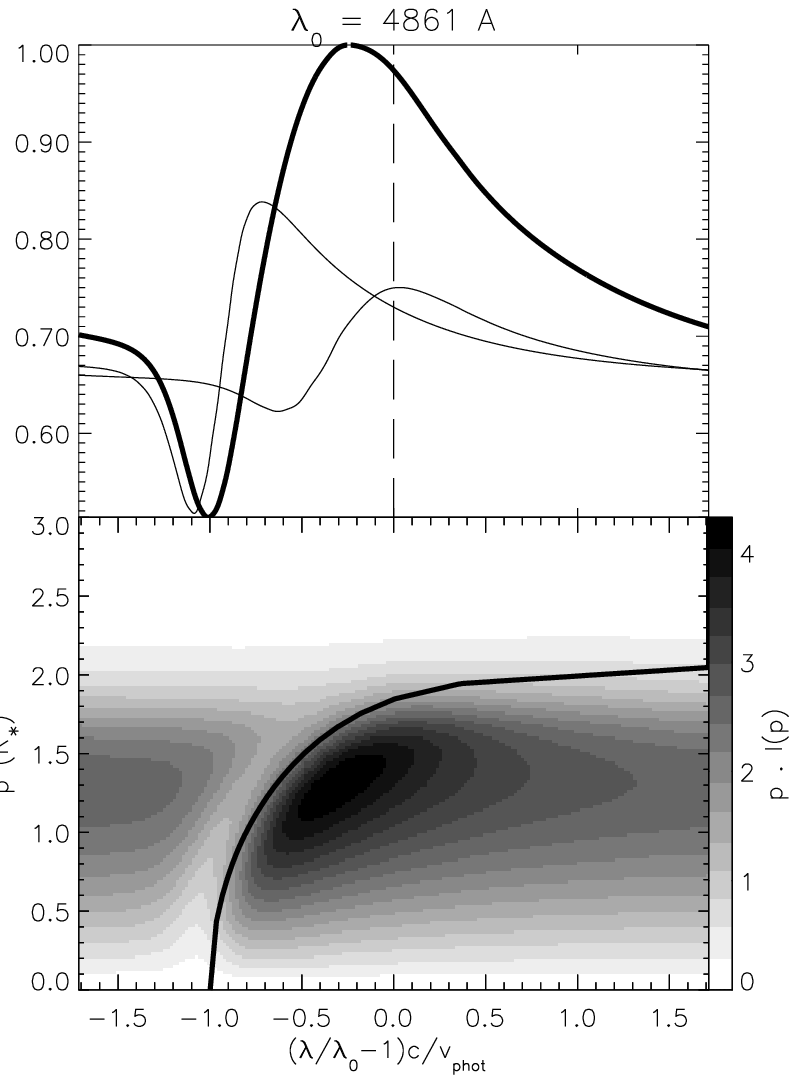, width=8cm}
\caption{Same as previous figure for H$\beta$.
Overplotted curves in the top panel correspond to rays with
$p=0.05 R_{\ast}$ and $p = 1.8 R_{\ast}$.
Note that the velocity location of the flux minimum coincides with $v_{\rm phot}$.
}
\label{fig_jnu2_hbeta}
\end{figure}

\begin{figure}[htp!]
\epsfig{file=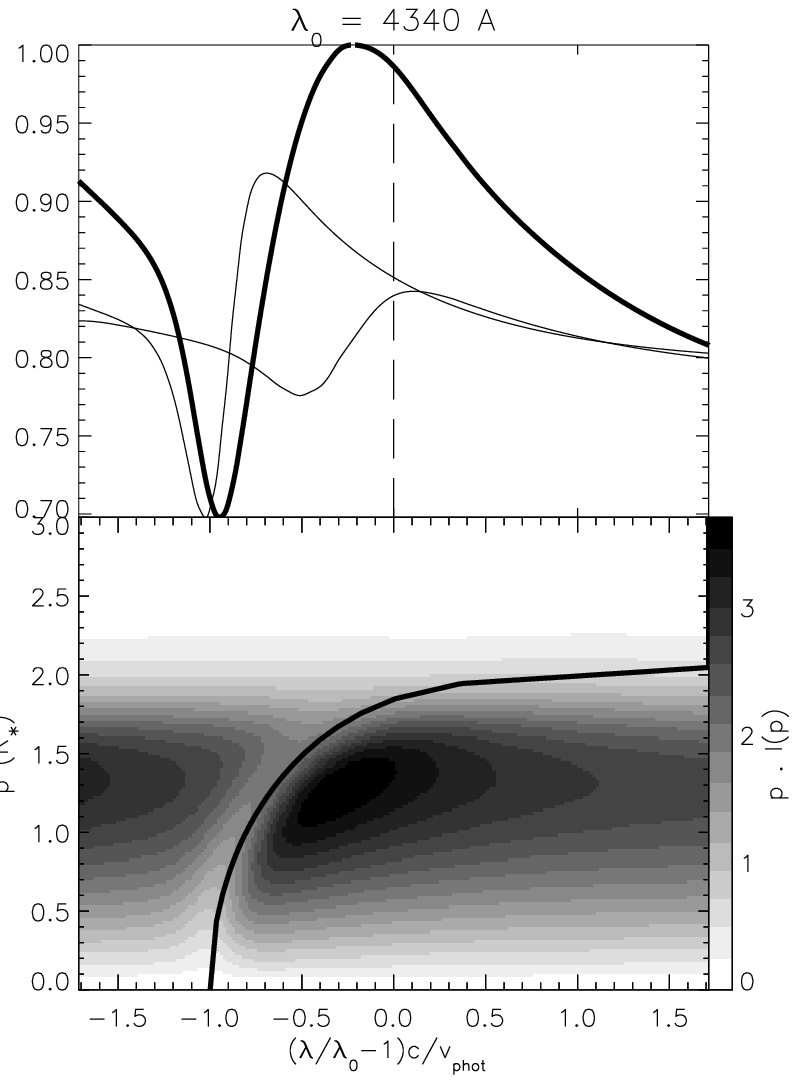, width=8cm}
\caption{Same as previous figure for H$\gamma$.
Overplotted curves in the top panel correspond to rays with
$p=0.05 R_{\ast}$ and $p = 1.6 R_{\ast}$.
Note that the velocity location of the flux minimum corresponds to a value
smaller than $v_{\rm phot}$.
}
\label{fig_jnu2_hgamma}
\end{figure}

In Figs.~\ref{fig_jnu2_hbeta}-\ref{fig_jnu2_hgamma},
we reproduce Fig.~\ref{fig_jnu2_halpha} for the cases of H$\beta$ and H$\gamma$. 
The correspondence with Fig.~\ref{fig_jnu2_halpha} is striking, only mitigated by the
difference in optical-depth (and line source functions) between these lines (Fig.~\ref{fig_tau}).
Moving from H$\alpha$ to H$\beta$ and H$\gamma$, the line optical depth decreases
and the velocity at maximum absorption in the line profile shifts towards line center,
i.e. from 1.1 to 1 and 0.95, in agreement with the values given above and
measured on the synthetic spectrum (Fig. 3, Paper I), which includes all bound-bound 
transitions of all species.

\begin{figure}[htp!]
\epsfig{file=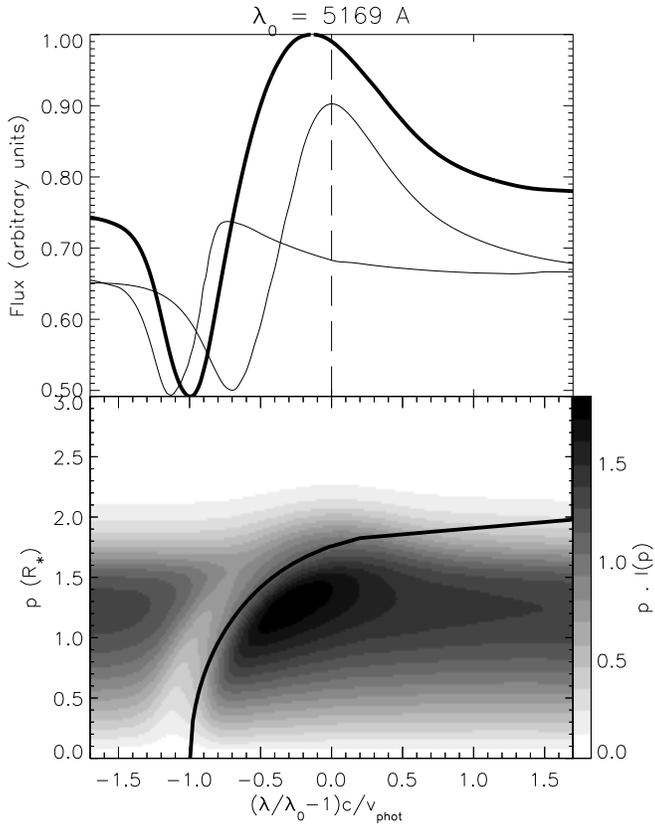, width=8cm}
\caption{Same as previous figure for Fe{\,\sc ii}5169\AA, in the cool model
presented in Sect. 3.3 of Paper I, corresponding to a late-stage model
in which Fe{\,\sc ii} lines are present throughout the V band.
For this plot, we reset the oscillator strength of all other Fe{\,\sc ii} contributors
to this wavelength region to essentially zero, in order to highlight the
formation region of the single transition at 5169\AA.
Overplotted curves in the top panel correspond to rays with
$p=0.04 R_{\ast}$ and $p = 1.5 R_{\ast}$.
Note that the velocity location of the flux minimum corresponds to a value
consistent with $v_{\rm phot}$, just like for H$\beta$ in 
Fig.~\ref{fig_jnu2_hbeta} for an intermediate-stage (hotter) model.
}
\label{fig_jnu2_fe2}
\end{figure}

For completeness, using the late-stage model of Sect. 3.3 (Paper I), we show in Fig.~\ref{fig_jnu2_fe2} 
the analogue of Fig.~\ref{fig_jnu2_halpha} for Fe{\sc ii}\,5169\AA\, 
(the oscillator strength of all bound-bound transitions of Fe{\sc ii} located within 500\AA\, 
of the line center were set to zero). 
The maximum absorption occurs at $|x| \sim 1$, and thus delivers exactly the photospheric 
velocity, in agreement with the velocity measurement of 0.99$v_{\rm phot}$ obtained from 
the synthetic spectrum feature attributed to that line (Fig. 5, Paper I), which as above
includes all bound-bound transitions of all species.

Detailed model atmosphere calculations of type II photospheric-phase SN can thus
deliver accurate and meaningful estimates of the photospheric velocity, by treating
accurately the complex continuum and line optical depth effects controlling the
line profile shape.
The agreement between measurements done on line profiles assuming either a single-line
or all possible lines contributions suggests that line-overlap does not affect noticeably
the inference of $v_{\rm phot}$ from the estimate of $v_{\rm abs}$ on the observed spectrum
of photospheric-phase type II SN.


\section{Conclusion}
\label{Sec_conc}

In this paper, we have presented and discussed various aspects of
the Expanding Photosphere Method, using knowledge drawn from a large 
ensemble of CMFGEN models for photospheric-phase type II SN.

We have obtained such CMFGEN-based correction factors from the minimisation
technique at the heart of the EPM, used to compensate for the 
approximation of the SN SED by a (unique-temperature) blackbody.
The correction factors do not just compensate for the effects of
dilution due to electron-scattering -- they also account
for the range of temperatures at the depth of thermalization, and,
most importantly, the presence of lines that alter considerably the SED at late times.
This latter fact, usually ignored by other groups, is what causes correction 
factors to take values of up to 1.5 in our grid of models,
while the upper limit for dilution factors is, by definition, unity.
Compared to previous tabulations of correction factors given
by E96, our values are systematically higher, by ca. 0.1., although the reason 
is unclear.

To gain further insight, we have extracted from these models the 
thermalization and photospheric radii. The ratio of these two radii
is related to the dilution factor.
We find that the ratio of the thermalization radius to the photospheric
radius is relatively uniform throughout the optical
and near-IR, and increases as the outflow cools down and hydrogen
recombines.
For a given outflow ionization or effective temperature, the ratio
is changed if the density exponent or the scale of
the supernovae is modified, resulting from the modulation of the
electron-scattering optical depth at the photosphere.
We also find that if one accounts for line opacity, the location
of the photosphere radius can vary by up to a factor of 2-3,
the effect being maximum in the UV and more modest in the optical and
beyond. Nonetheless, this reveals that the uniqueness of the
photosphere radius for all wavelengths holds better at earlier times,
when type II SN spectra are more featureless, closer to the equivalent
continuum energy distribution.
This has some bearing on the adequacy of using the Baade method for
distance determinations.

In addition to the determination of correction factors,
the accuracy of the measurement of the photospheric velocity is equally
important for the reliability of the EPM.
We have shown that, contrary to expectations, optically thick lines
do not necessarily show a minimum P-Cygni line profile flux at a
Doppler-shifted wavelength that corresponds to the photospheric velocity.
Instead, depending on the outflow properties, such a measurement can
deliver an overestimate or an underestimate of the photospheric
velocity.
This is particularly problematic for earlier models which show broad
P-Cygni line profile troughs, mostly for hydrogen Balmer lines.
Unfortunately we have also demonstrated that, due to the 
more well defined photospheric radius, the
lack of contaminating lines and a SED closer to that of a blackbody,
it is at these earlier times that the EPM is best used.

This paper has established some foundations for a careful use
of the EPM, which we are currently applying to SN1999em.

\begin{acknowledgements}

We wish to thank Mario Hamuy for assistance with the photometry.
DJH gratefully acknowledges partial support for this work from NASA-LTSA 
grant NAG5-8211.  

\end{acknowledgements}

\end{document}